\documentclass[acmtog, dvipsnames,table,xcdraw]{acmart}

\usepackage{booktabs} %
\usepackage{todonotes}
\usepackage[overload]{empheq} %
\usepackage{algorithm}
\usepackage{algpseudocode}
\usepackage{amsmath}
\usepackage{xcolor}

\usepackage{multirow}
\usepackage{graphicx}

\setcopyright{rightsretained}
\acmJournal{TOG}
\acmYear{2022} \acmVolume{1} \acmNumber{1} \acmArticle{1} \acmMonth{1} \acmPrice{}\acmDOI{10.1145/3527660}

\begin{document}
\title{DiffCloth: Differentiable Cloth Simulation with Dry Frictional Contact}

\author{Yifei Li}
\affiliation{%
 \institution{MIT CSAIL}
 \city{Cambridge}
 \state{MA}
 \country{USA}
 }
\email{liyifei@csail.mit.edu}

\author{Tao Du}
\affiliation{%
 \institution{MIT CSAIL}
 \city{Cambridge}
 \state{MA}
 \country{USA}
 }
\email{taodu@csail.mit.edu}

\author{Kui Wu}
\affiliation{%
 \institution{Tencent Lightspeed \& Quantum Studios}
 \country{USA}
 }
\email{kwwu@tencent.com}

\author{Jie Xu}
\affiliation{%
 \institution{MIT CSAIL}
 \city{Cambridge}
 \state{MA}
 \country{USA}
 }
\email{jiex@csail.mit.edu}

\author{Wojciech Matusik}
\affiliation{%
 \institution{MIT CSAIL}
 \city{Cambridge}
 \state{MA}
 \country{USA}
 }
\email{wojciech@csail.mit.edu}

\renewcommand\shortauthors{Li. et al}

\newcommand{\x}[0]{\mathbf{x}}
\newcommand{\y}[0]{\mathbf{y}}
\newcommand{\z}[0]{\mathbf{z}}
\newcommand{\n}[0]{\mathbf{n}}

\newcommand{\p}[0]{\mathbf{p}}
\newcommand{\rforce}[0]{\textbf{r}}
\newcommand{\rimpulse}[0]{\textbf{r}}
\newcommand{\uu}[0]{\mathbf{u}}
\newcommand{\vecv}[0]{\mathbf{v}}
\newcommand{\vecf}[0]{\mathbf{f}}
\newcommand{\vecd}[0]{\mathbf{d}}
\newcommand{\vel}[0]{\mathbf{v}}
\newcommand{\M}[0]{\mathbf{M}}
\newcommand{\fex}[0]{\mathbf{f}_\text{ext}}
\newcommand{\fin}[0]{\mathbf{f}_\text{int}}
\newcommand{\sn}[0]{ \mathbf{s}_{n}}
\newcommand{\bvec}[0]{\mathbf{b}}
 
\newcommand{\W}[0]{\mathbf{W}}

\newcommand{\R}[0]{\mathbb{R}}

\newcommand{\A}[0]{\mathbf{A}}
\newcommand{\B}[0]{\mathbf{B}}
\newcommand{\J}[0]{\mathbf{J}}
\newcommand{\Con}[0]{\mathcal{C}}

\newcommand{\yl}[1]{\textcolor{teal}{[\colorbox{teal}{\textcolor{white}{\textbf{YIFEI}}}: #1]}}

\newcommand{\hlred}[1]{ #1 } %
\newcommand{\hltodo}[1]{ \textcolor{regulation_red}{#1} } %
\newcommand{\hlrevision}[1]{#1}

\newcommand{\kw}[1]{\textcolor{orange}{[\textbf{KUI}: #1]}}
\newcommand{\td}[1]{\textcolor{muted_navy_blue}{[[\colorbox{muted_navy_blue}{\textbf{\textcolor{white}{TAO}}}: #1]}}
\newcommand{\jx}[1]{\textcolor{cyan}{[\textbf{JIE}: #1]}}
\newcommand{\wm}[1]{\textcolor{yellow}{[\textbf{WOJCIECH}: #1]}}

\newcommand{\ContactSet}[0]{\mathcal{I}}

\newcommand{\Pmat}[0]{\mathbf{P}}
\newcommand{\Cmat}[0]{\mathbf{C}}
\newcommand{\Rot}[0]{\mathbf{R}}
\newcommand{\Rmat}[0]{\mathbf{R}}
\newcommand{\Mmat}[0]{\mathbf{M}}
\newcommand{\Imat}[0]{\mathbf{I}}
\newcommand{\diffpd}[0]{\textit{DiffPD}}
\newcommand{\dd}[0]{\mathbf{d}}
\newcommand{\tb}[0]{\mathbf{t}}

\newcommand{\norm}[1]{\Vert #1 \Vert} %

\definecolor{peach}{rgb}{ 0.943, 0.188, 0.526}
\definecolor{plum}{rgb}{ 0.858, 0.188, 0.478}
\definecolor{muted_navy_blue}{RGB}{63, 75, 166}
\definecolor{muted_sky_blue}{RGB}{134,166,213}
\definecolor{federal_blue}{RGB}{0,96,240}
\definecolor{regulation_red}{RGB}{226, 20, 79}
\definecolor{federal_gold}{RGB}{240, 212, 14}

\newcommand{\deprecated}[1]{}

\begin{abstract}

Cloth simulation has wide applications in computer animation, garment design, and robot-assisted dressing. This work presents a differentiable cloth simulator whose additional gradient information facilitates cloth-related applications. Our differentiable simulator extends a state-of-the-art cloth simulator based on Projective Dynamics (PD) and with dry frictional contact~\cite{Ly20_PDdryFriction}. We draw inspiration from previous work~\cite{Du21_DiffPD} to propose a fast and novel method for deriving gradients in PD-based cloth simulation with dry frictional contact. Furthermore, we conduct a comprehensive analysis and evaluation of the usefulness of gradients in contact-rich cloth simulation. Finally, we demonstrate the efficacy of our simulator in a number of downstream applications, including system identification, trajectory optimization for assisted dressing, closed-loop control, inverse design, and real-to-sim transfer. We observe a substantial speedup obtained from using our gradient information in solving most of these applications.

\end{abstract}



\begin{teaserfigure}
    \centering
    \includegraphics[width=\textwidth]{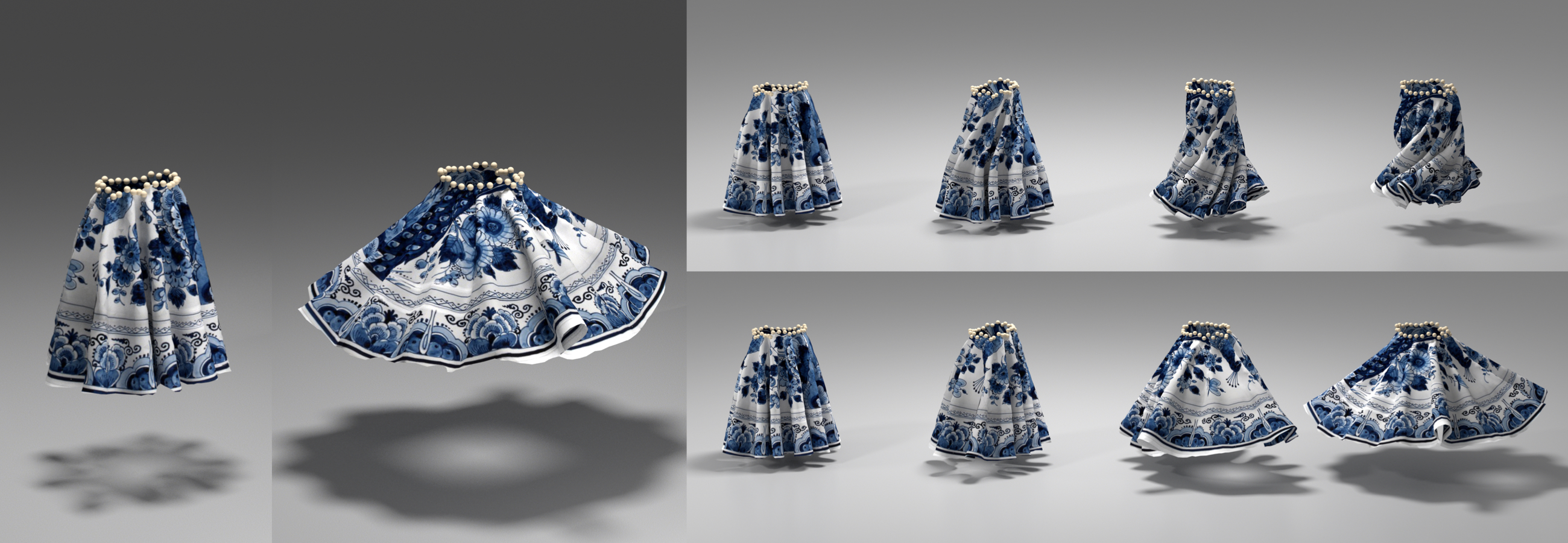}
    \caption{
    We present a differentiable cloth simulator with dry frictional contact and demonstrate its efficacy in multiple downstream applications, including designing this twirl dress (10902 degrees of freedom and 125 time steps with a step size of $1/120$ seconds). The goal is to optimize the material parameters of the dress so that the apex angle of the cone it forms after a twirl reaches a desired value (100 degrees in this example). Left and middle: the optimized dress before and after a twirl. Top right and bottom right: motion sequences of the twirl dress before and after material parameter optimization using our differentiable simulator. The final apex angles before and after optimization are 39.06  and 100.30 degrees, respectively.
    }
    \label{fig:teaser}
\end{teaserfigure}

\maketitle

\section{Introduction}\label{sec:intro}

Clothing is ubiquitous in our daily lives. With the widespread appearance of clothing in the fashion industry, film industry, computer animation, and video games, simulating cloth has been an active research topic for more than two decades. Today, research advancement in cloth simulation has unlocked various applications such as virtual try-on~\cite{Guan2012DRAPE}, garment design~\cite{Umetani2011SensitiveCouture, Wang2018RuleFree, Bartle2016PatternAdjustment, Montes2020tight}, fold design~\cite{li2018Foldsketch}, garment grading~\cite{Brouet2012Transfer}, sagging-free inversion~\cite{Ly2018inverse}, and robot-assisted dressing~\cite{Clegg2020RobotAssistedDressing, Clegg2018LearningtoDress}.

Inspired by the recent development of differentiable physics simulation and its success in rigid-body systems~\cite{De18_End2End, Geilinger20_ADD}, fluidic systems~\cite{hu2019difftaichi, Du20_Stokes}, and deformable-body systems~\cite{Hahn19_Real2Sim, Geilinger20_ADD}, we argue in this paper that a large number of cloth-related applications would also benefit from a high-quality differentiable cloth simulator. The critical ingredient in previous differentiable simulators is their ability to compute gradients by backpropagating any differentiable performance metrics through simulation. Such additional gradient information unlocks gradient-based continuous optimization methods, which often bring a substantial speedup compared with their gradient-free counterparts in downstream applications.

Compared with the recent active development of differentiable simulators for rigid-body and soft-body dynamics, research work about differentiable cloth simulation is still relatively sparse. Indeed, cloth simulation introduces unique challenges from its co-dimensional dynamics and, in particular, rich contact events. While many differentiable simulators have provided solutions to derive gradients for contact models of varying complexity, their techniques typically do not expect contact to be as frequent as in cloth simulation. Deriving gradients with frequent contact and self-collisions in cloth simulation is not fully resolved even in the state-of-the-art differentiable cloth simulators~\cite{Liang19_DiffCloth}, and our work attempts to fill this gap.

In this work, we present a differentiable cloth simulator with extra care of its contact model. We base our method on the state-of-the-art cloth simulator from~\citet{Ly20_PDdryFriction} and employ its Projective Dynamics (PD) simulation method and dry frictional contact described by the Signorini-Coulomb law. Therefore, our differentiable cloth simulator inherits both the speedup from Projective Dynamics and the physical accuracy from the dry frictional contact model. Unlike previous papers relying on automatic differentiation tools to derive gradients, we present an iterative solver modified from~\citet{Du21_DiffPD} to accommodate the dry frictional contact model. We show that the modified iterative solver leads to a substantial speedup over a standard linear solver in gradient computation.

To have well-defined gradients, a differentiable simulator expects all quantities computed in simulation to be sufficiently smooth. However, the non-smooth position and force changes from large numbers of contact events question this fundamental assumption. To fully understand the usefulness of a differentiable simulator in contact-rich environments, our work conducts comprehensive evaluation and analysis of the behavior of our differentiable cloth simulation with a varying number of contact events. While previous papers have provided similar discussions, they primarily focus on understanding penalty-based contact models~\cite{Geilinger20_ADD} or discussing collisions in rigid-body dynamics~\cite{Werling-RSS-21} where contact events are not as frequent as in cloth simulation. To our best knowledge, our work is the first to present such evaluation and discussion of the usefulness of gradients in a differentiable cloth simulator with dry frictional contact.

We demonstrate the efficacy of our simulator in various applications, including system identification of frictional coefficients in cloth simulation, inverse garment design for computer animation, and motion planning of robotic manipulators in robot-assisted dressing. Many of these applications would either be impossible or require a much longer time to solve with existing methods. With the extra gradient information from our differentiable simulator, we unlock gradient-based optimizers to solve these problems with a much higher sample efficiency than traditional gradient-free methods.

To summarize, our work contributes the following:
\begin{itemize}
    \item We present a novel differentiable cloth simulator with dry frictional contact and an iterative solver for speeding up its gradient computation.
    \item We evaluate the source of non-differentiability in the dry frictional contact model and discuss the usefulness of gradients in differentiable, contact-rich cloth simulation.
    \item We show the efficacy of our simulator in various applications, including system identification, trajectory optimization for assisted dressing, closed-loop control, inverse design, and real-to-sim transfer.
\end{itemize}

\section{Related Work}\label{sec:related}
Our work is closely related to cloth simulation and its applications in computer graphics. It is also relevant to the more recent differentiable simulation methods developed in the graphics and machine learning communities.

\paragraph{Cloth simulation}
Physics-based cloth simulation has been a popular topic in the graphics community for decades since clothing has been widely used in our daily lives. The implicit Euler integration is used to simulate cloth robustly with large time steps~\cite{Terzopoulos1987Elastically, Baraff1998cloth} while introducing excessive numerical damping. To alleviate this problem, implicit and explicit methods (IMEX)~\cite{Bridson2003MixedIntegration, Stern2009IMEX} explicitly integrate elastic forces and implicitly integrate damping forces. Researchers have also introduced other variational integrators, e.g., BDF2~\cite{Choi2002BDF2} and Symplectic integrator~\cite{Stern2006Symplectic}, to conserve the total energy of the system.

For cloth modeled as a mass-spring system, \citet{Liu2013FastMassSpring} treat the implicit Euler integration as an energy minimization problem. The global linear system then remains constant on run-time, and each spring constraint can be solved separately in a local step. That global/local solver idea is generalized as Projective Dynamics (PD)~\cite{Bouaziz14_PD}, which supports material models whose elastic energy has a specific quadratic form.  Projective Dynamic is further extended to support more general materials~\cite{overby2017admmpd, Liu17_Quasi}. Though a local step in PD can be processed in parallel, the global step still needs to maintain a large pre-factorized sparse matrix and do back substitutions in each step. Another relevant idea is Position-Based Dynamics (PBD)~\cite{Muller2007PBD, Macklin2016XPBD}, which iteratively projects each constraint in a non-linear Gauss-Seidel-like fashion, leading to highly parallelizable computation on GPUs.

Both PD and PBD have led to a few follow-up works on more advanced speedup techniques. \citet{Komaritzan18_projectiveSkinning} present techniques to speed up PD with contact for physics-based character skinning. For PD and PBD in cloth simulation, \citet{Wang2015Chebyshev} and a follow-up work \cite{wang2016Descent} propose a Chebyshev acceleration technique that can be applied to both PD and PBD to speed up convergence. \citet{Fratarcangeli2016Vivace} also introduce a parallel randomized Gauss-Seidel method that re-organizes the unknowns of the sparse linear system into a few independent blocks, which can be solved in parallel with a single Gauss-Seidel step. Recently, the computation of integration is further accelerated by the multigrid method, e.g., geometric multigrid scheme~\cite{wang2018Multigrid}, algebraic multigrid scheme~\cite{Tamstorf2015Multigrid}, and Galerkin multigrid scheme~\cite{Xian2019Galerkin}.

Lastly, our work is also relevant to previous attempts to matching cloth simulation with real fabric behaviors~\cite{Miguel2012DataDriven, Wang2011DataDriven,Miguel2013DataDriven,Clyde2017DataDriven}, which is typically done by fitting constitutive material models with real material properties and can benefit from extra gradient information from a differentiable simulator~\cite{Hahn19_Real2Sim}.

\paragraph{Cloth contact and friction} Contact and friction are key ingredients in modern cloth simulation. \citet{Provot1997ImpactZone} proposes a method called ``impact zones'' to collect the nodes involved in multiple collisions into impact zones, which are treated as rigid bodies.
\citet{Harmon2008ImpactZone} improve the failsafe of impact zones by allowing some sliding motion of the incriminated vertices. \citet{Bridson2002RobustCollisions} introduce a hybrid method to combine the idea of applying repulsion impulses, frictions, and impact zones to handle cloth collision robustly, which is still widely used in modern cloth simulators~\cite{Narain2012arcsim, Li2020pcloth}. To model friction fully implicitly, \citet{Brown2018Dissipative} treat friction as an additional dissipative term in optimization. Regarding contact handling, repulsive forces or penalty methods have been widely used~\cite{Bouaziz14_PD, Wang2015Chebyshev, Geilinger20_ADD, Macklin2020PrimalDual} since they are easy to implement. However, these methods often need high stiffness in the penalty energy, leading to disturbing jittering artifacts that demand careful tuning. Recently, \citet{Li2020IPC, Li2021CIPC} define a smooth dissipative potential for friction using barrier methods. Their method can guarantee interpenetration-free states after each time step without the need for parameter tuning.

Unlike penalty-based methods, constraint-based collision handling methods formulate contact as constraints in the physics system. \citet{Otaduy2009ImplicitContact} first formulate cloth contacts as a sparse linear complementarity problem (LCP). Their key idea is to interleave frictional contact iterations with normal contact iterations. Instead of using a pyramidal approximation of the Coulomb friction cone~\cite{Otaduy2009ImplicitContact}, \citet{li2018ImplicitFrictional} rely upon the exact Coulomb friction cone and adaptively refine nodes to ensure an accurate treatment of frictional contact. Although constraint-based methods typically ensure the physics-based constraints characterizing contact and friction are satisfied after time integration, their computation cost is typically much more expensive than a penalty-based method. Recently, \citet{Ly20_PDdryFriction} propose an efficient algorithm to incorporate frictional contact into Projective Dynamics so that non-penetration and Coulomb constraints are satisfied simultaneously in a semi-implicit way.

\paragraph{Inverse dynamics}
Inverse dynamics have been studied in robotics for decades to reconstruct internal forces or torques from the observations of robotic systems. However, existing methods usually focus on rigid-body systems only, which have less than a hundred degrees of freedom (DoFs)~\cite{5509646, 7803330, kang2021animal}. Inverse dynamics for high-DoF systems like soft bodies, fluids, and cloth are still under exploration due to the lack of high-quality numerical solutions in robotics for both simulation and differentiation. One noticeable distinction between inverse dynamics and differentiable simulation is that differentiable simulation computes additional gradients for initial states, system parameters, and design parameters. Therefore, differentiable simulation enables more applications like system identification, inverse design, and real-to-sim matching that traditional inverse dynamics typically do not consider.

\paragraph{Differentiable simulation}
Differentiable simulation is a relatively recent concept explored in the graphics and machine learning community, but its original idea can be traced back to much earlier works in graphics decades ago. Perhaps one of the earliest such papers is~\citet{Witkin88_spaceTime}, which shows optimizing simulation to minimize an objective. Despite the recent advances in differentiable simulators in rigid-body dynamics~\cite{popovic2003motion, de2018end,toussaint2018differentiable,degrave2019differentiable,qiao2020scalable, Xu-RSS-21}, soft-body dynamics~\cite{Hu19_ChainQueen,hu2019difftaichi,Hahn19_Real2Sim, Geilinger20_ADD,Du21_DiffPD}, and fluid dynamics~\cite{treuille2003keyframe, mcnamara2004fluid, wojtan2006keyframe, schenck2018spnets,holl2020learning}, differentiable cloth simulation still lacks a good solution. One natural idea is to use particle-based strategies that approximate a nodal system of cloth with graph neural networks~\cite{li2018learning,sanchez2020learning}. Although the neural networks are naturally differentiable, the physical accuracy is hard to guarantee.

To accurately predict the behavior of real-world objects, recent papers~\cite{Hahn19_Real2Sim, Geilinger20_ADD} present differentiable soft-body systems and mass-spring systems with implicit Euler time integration and penalty-based contacts. Although a few recent papers~\cite{Macklin2020PrimalDual, Li2020IPC, Li2021CIPC} have introduced differentiable contact handling methods, none of them show a clothing example using the differentiability besides demonstrating the differentiability of their methods in theory. \citet{Liang19_DiffCloth} are the first to introduce a fully functional differentiable cloth simulation with contact, friction, and self-collision. Instead of constructing a static LCP problem, they develop a quadratic programming (QP) problem to minimize the change between the collision-free state and the original mesh state. \citet{murthy2021gradsim} also present a differentiable cloth simulator with penalty-based frictional contact in their fully differentiable simulation and rendering pipeline. In this paper, we build upon the differentiable PD framework~\cite{Du21_DiffPD} to simulate cloth dynamics with dry frictional contact~\cite{Ly20_PDdryFriction} and augment it with gradient computation.

A significant challenge in differentiable simulation is the presence of non-smooth contact events, where non-smoothness can arise from discontinuous contact shapes~\cite{Popovic00_Manipulation}, impulsive forces~\cite{Hu19_ChainQueen}, or branches in contact laws~\cite{Ly20_PDdryFriction}, to name a few. Many existing differentiable simulators assume such non-smoothness does not affect the usability of gradients. Indeed, they present empirical results that observe the benefits of using gradients in downstream applications~\cite{De18_End2End,Du21_DiffPD,Liang19_DiffCloth,Geilinger20_ADD}. While our work also observes the advantages of gradients over gradient-free methods in several applications, we take one step further by analyzing the source of non-smoothness and discontinuities in contact-rich cloth simulation. We hope that our experience can serve as useful heuristics for applying differentiable simulation methods in the future.
\section{Background}\label{sec:background}

In this section, we briefly review the Projective Dynamics cloth simulation method with dry frictional contact described in~\citet{Ly20_PDdryFriction}, on which our differentiable cloth simulator is based. Our core contribution lies in the development of gradient computation in this PD framework with dry frictional contact, which we detail in the next section.

\subsection{Projective Dynamics Method}

\paragraph{Implicit time integration} We model cloth as a nodal system with $m$ 3D nodes. Let $\x(t)$ and $\vel(t)$ be two vector functions from $\R^+$ to $\R^{3m}$ indicating the positions and velocities of all nodes at time $t$. We consider the standard implicit time-stepping scheme discretizing $\x(t)$ and $\vel(t)$ as follows:
\begin{align}
    \x_{n+1} = &\x_n + h\vel_{n+1}, \label{eqn:implicit_x}\\
    \vel_{n+1} = &\vel_n + h\M^{-1}[\fin(\x_{n+1}) + \fex], \label{eqn:implicit_v}
\end{align}
where $h$ is the time step size, $\M$ a positive diagonal mass matrix of size $3m\times 3m$, and $\fin$ and $\fex$ the internal and external forces at each node stacked as two $3m$-dimensional vectors, respectively. Note that we assume for now $\fin$ and $\fex$ do not contain contact forces, which will be discussed separately in Sec.~\ref{sec:background:contact}. We use the cloth material model described in~\citet{Bouaziz14_PD} to define $\fin$. The implicit time-stepping scheme connects the state of the nodal system $(\x_n, \vel_n)$ at time $t_n$ to $(\x_{n+1}, \vel_{n+1})$ at new time $t_{n+1}=t_n+h$.

\paragraph{Optimization view} As discussed by~\citet{Martin11_ElasticMaterial}, the implicit time-stepping scheme can be rephrased as finding a saddle point of the following energy minimization problem:
\begin{align}
    \min_{\x_{n+1}}\; \underbrace{\frac{1}{2h^2}(\x_{n+1} - \y)^\top\M(\x_{n+1} - \y)+W(\x_{n+1})}_{g(\x_{n+1})}, \label{eqn:implicit_energy}
\end{align}
where $\y = \x_{n}+h \vel_n + h^2 \M^{-1} \fex$ is a constant vector that can be precomputed at the beginning of the current time step. The potential energy $W$ defines the internal force $\fin$ by its spatial gradients: $\fin = -\nabla W(\x_{n+1})$. It is easy to verify that setting the gradient of the objective function to zero leads to a system of equations identical to Eqns. (\ref{eqn:implicit_x}) and (\ref{eqn:implicit_v}).

\paragraph{Local and global solvers in PD} The key assumptions in PD is that the internal energy $W$ can be written as a sum of quadratic forms~\cite{Bouaziz14_PD,Liu17_Quasi}:
\begin{align}
    W_i(\x) = & \min_{\p_i\in\mathcal{M}_i}\;\underbrace{\frac{w_i}{2}\|\A_i\x - \p_i\|_2^2}_{\widetilde{W}_i(\x, \p_i)}, \label{eqn:pd_project} \\
    W(\x) =& \sum_i W_i(\x). \label{eqn:pd_energy}
\end{align}
Here, each energy $W_i$ projects $\A_i\x$, a linear transformation of $\x$, to its closest point in the set $\mathcal{M}_i$ and scales its squared distance by a prespecified stiffness $w_i$. Both $\A_i$ and $\mathcal{M}_i$ are predefined and independent of $\x$. Here, $\mathcal{M}_i$ is the constraint manifold, and $\p_i$ is the auxiliary projection variable as defined in~\citet{Bouaziz14_PD}.

With such an assumption on $W$, PD proposes a local-global solver to minimize a surrogate objective function:
\begin{align}
    \widetilde{g}(\x_{n+1},\p)=\frac{1}{2h^2}(\x_{n+1}-\y)^\top\M(\x_{n+1}-\y)+\sum_i \widetilde{W}_i(\x_{n+1},\p_i),
\end{align}
where $\p$ stacks up all $\p_i$ from each $W_i$. In the local step, PD fixes $\x_{n+1}$ and projects each $\p_i$ to its corresponding $\mathcal{M}_i$ by solving Eqn. (\ref{eqn:pd_project}). Such a local step can be done in parallel for each $\widetilde{W}_i$. In the global step, PD fixes $\p$ and minimizes $\widetilde{g}$ as a function of $\x_{n+1}$, which turns out to have a closed-form solution:
\begin{align}
\underbrace{(\M+h^2\sum_i w_i \A_i^\top\A_i)}_{\Pmat}\x_{n+1} = \underbrace{\M\y + h^2\sum_i w_i \A_i^\top\p_i}_{\bvec(\p)}. \label{eqn:pd_global}
\end{align}
By alternating between the local and global steps, PD monotonically decreases the surrogate energy $\widetilde{g}$ until convergence, which can be shown to agree with the saddle point of $g$~\cite{Liu17_Quasi}. The source of efficiency in PD comes from the observation that $\Pmat$ is a constant matrix that can be prefactorized at the beginning of the simulation, leading to an efficient global step requiring back-substitution only.

\subsection{Dry Frictional Contact in Projective Dynamics}\label{sec:background:contact}

\paragraph{Signorini-Coulomb law} \citet{Ly20_PDdryFriction} augment the standard PD framework described above with non-penetration collision and Coulomb friction by assuming contact applies to nodes only. At each time step, assuming a collision detection algorithm has identified a set $\ContactSet\subseteq\{1,2,3,\cdots,m\}$ which describes the indices of contact nodes. For each node $j\in\ContactSet$, the Signorini-Coulomb law~\cite{Brogliato16_NonsmoothLagSys} requires its local force $\rforce_j$ and velocity $\uu_j$ to satisfy one of the following three conditions:
\begin{subequations}\label{eqn:contact_all}
\begin{align}[left = \empheqlbrace\,]
&\text{Take off: } \rforce_j = \mathbf{0}, \uu_{j|N} > 0, \label{eqn:contact_1} \\
&\text{Stick: }  \norm{\rforce_{j|T}} < \mu \rforce_{j|N}, \uu_j=\mathbf{0}, \label{eqn:contact_2}\\
  &\text{Slip: } \norm{\rforce_{j|T}} = \mu \rforce_{j|N},  \uu_{j|N}=0,  \rforce_{j|T} \parallel \uu_{j|T}, \rforce_{j|T} \cdot \uu_{j|T} \leq 0.  \label{eqn:contact_3}
\end{align}
\end{subequations}
Here, $\mu$ is the frictional coefficient, and $\uu_j$ and $\rforce_j$ are the nodal velocity and contact force represented in the local contact frame spanned by the tangential plane and contact normal on the contact surface. The notations $j|T$ and $j|N$ represent the tangential and normal components of a 3D vector defined in the local frame. We further define $\Con^j\subseteq\R^3\times\R^3$ as the set of valid $(\rforce_j, \uu_j)$ pairs described by the conditions above, allowing us to rewrite Eqn.~\ref{eqn:contact_all} compactly: $(\rforce_j, \uu_j)\in\Con^j$. 

\paragraph{Implicit time integration with dry frictional contact} The original implicit time integration now needs to be augmented with additional constraints describing contact conditions~\cite{Ly20_PDdryFriction}:
\begin{equation}\label{eqn:implicit_with_contact}
   \left\{\begin{array}{lr}
    \x_{n+1}   = \x_{n} + h\vel_{n+1}, \\
    \vel_{n+1} = \vel_{n} + h\mathbf{M}^{-1}[\fin(\x_{n+1})+\fex +\J_n^\top(\x_{n}, \vel_{n}) \rforce_{n+1}], \\
    \uu_{n+1} = \J_n(\x_{n}, \vel_{n}) \vel_{n+1}, \\
    (\rforce_{n+1,j}, \uu_{n+1,j}) \in \Con^j_{n},   \forall j\in\ContactSet_n.
     \end{array}\right.
\end{equation}
Here, we have added the subscript $n$ in the definitions of the contact set $\ContactSet$ and the contact condition $\Con^j$ described above to specify the time step they are defined from. The notation $\rforce_{n+1,j}$ and $\uu_{n+1,j}$ select the 3D force and velocity corresponding to the $j$-th node from $\rforce_{n+1}$ and $\uu_{n+1}$, respectively. The Jacobian matrix $\J_n$ of size $3 |\ContactSet_n|\times 3m$ selects global vectors defined on the contact nodes and computes their coordinates in the local contact frames. Note that our definition of $\J_n$ implies that it is computed in an explicit manner, i.e., it relies on the last state $(\x_n, \vel_n)$ entering the $n$-th time step. For brevity, this background section will describe simple contact events between a node and a static plane, in which case $\J_n$ is a constant matrix that does not need to be computed from $(\x_n, \vel_n)$. Extensions to more sophisticated contact events, e.g., self-collisions between multiple contact nodes or contact events between a node and a non-planar obstacle, can be found in~\citet{Ly20_PDdryFriction} and in our implementation and experiments.

\paragraph{Projective Dynamics with dry frictional contact} The core idea in~\citet{Ly20_PDdryFriction} is an additional local step in PD that solves each contact node independently. Noting that the contact conditions are primarily defined on nodal velocities instead of nodal positions,~\citet{Ly20_PDdryFriction} first rewrite the global step in Eqn. (\ref{eqn:pd_global}) as an equation of velocities:
\begin{align}\label{eqn:pd_velocity}
    \Pmat\vel_{n+1} = \underbrace{\frac{1}{h}[\bvec(\p) - \Pmat \x_{n}]}_{\widehat{\bvec}(\p)}.
\end{align}
The velocity-based PD global-local steps then alternate between this new global solver and the original local step, which is unaffected by this change of variables. By comparing Eqn. (\ref{eqn:implicit_with_contact}) and Eqn. (\ref{eqn:pd_velocity}), we can see that incorporating the contact force adds an additional impulse $h\J_n^\top\rforce_{n+1}$ to the right-hand side of the global step:
\begin{align}\label{eqn:pd_velocity_with_contact}
    \Pmat \vel_{n+1}=\widehat{\bvec}(\p)+h\J_n^\top\rforce_{n+1}.
\end{align}
The goal of the global solver is now to find $(\vel_{n+1}, \rforce_{n+1})$ that satisfy contact conditions at each $j$. Noting that in Eqn. (\ref{eqn:pd_velocity_with_contact}), $\Pmat$ is the only operator that couples unknown contact forces from different $j$, ~\citet{Ly20_PDdryFriction} propose the following iterative solver based on the decomposition of $\Pmat$ to solve Eqn. (\ref{eqn:pd_velocity_with_contact}):
\begin{align}\label{eqn:ly_et_al_govrningEq}
    \M\vel_{n+1}^{k+1} = \widehat{\bvec}(\p) - \underbrace{(h^2\sum_{i} w_i\A_i^\top\A_i)}_{\Pmat - \M}\vel_{n+1}^{k} + h\J_n^\top\rforce_{n+1}^{k+1},
\end{align}
where the superscripts $k$ and $k+1$ indicate two consecutive iterations in the iterative solver at fixed time step $n+1$. After iteration $k$, $\rforce_{n+1}^{k+1}$ is updated using $\vel_{n+1}^{k}$ to enforce the Signorini-Coulomb law, which is then used to update $\vel_{n+1}^{k+1}$ according to Eqn.~(\ref{eqn:ly_et_al_govrningEq}).  It is easy to see that the proposed iterative solver fully decouples the momentum equation on each node, allowing~\citet{Ly20_PDdryFriction} to enforce the Signorini-Columb law by adjusting $(\vel_{n+1,j},\rforce_{n+1,j})$ on each contact node $j$ independently in a straightforward manner. When the iterative solver converges,~\citet{Ly20_PDdryFriction} prove that the result is indeed a solution to Eqn. (\ref{eqn:implicit_with_contact}). We refer the readers to the original paper for the details.

\section{Differentiable Cloth Simulation}\label{sec:method}

We now describe in detail how we extend the forward simulator in Sec.~\ref{sec:background} to build a differentiable cloth simulator. We start by differentiating implicit time integration in PD without contact, followed by explaining how contact gradients can be added to this framework. Compared with other differentiable simulators, our simulator is unique because of its treatment of the contact-rich nature of cloth simulation: many existing differentiable simulators focus on sparse contact events in rigid-body or deformable-body dynamics~\cite{Geilinger20_ADD,Hu19_ChainQueen,Du21_DiffPD}, and the state-of-the-art differentiable cloth simulation method~\cite{Liang19_DiffCloth} handles rich contact events but without physics-based contact forces or friction. To our best knowledge, our work is the first to present a differentiable cloth simulator that can handle rich contact events with Coulomb's law of friction.

\subsection{Gradients without Contact}

\paragraph{Differentiating implicit time integration} The core step in building a differentiable simulator based on implicit time-stepping scheme is to backpropgate gradients through the implicit integration described in Eqns. (\ref{eqn:implicit_x}) and (\ref{eqn:implicit_v}), or equivalently, to derive the Jacobian of the output $(\x_{n+1},\vel_{n+1})$ with respect to the input $(\x_n, \vel_{n})$. Mature techniques such as sensitivity analysis, adjoint method, and implicit function theorem have proven to be successful in computing such gradients~\cite{Hahn19_Real2Sim,Geilinger20_ADD,Du21_DiffPD}. To sketch the idea, we plug $\vel_{n+1}$ from Eqn. (\ref{eqn:implicit_v}) into Eqn. (\ref{eqn:implicit_x}):
\begin{equation}\label{eqn:backprop_x}
\begin{aligned}
    \x_{n+1} = &\; \x_n+h\vel_n+h^2\M^{-1}[\fin(\x_{n+1})+\fex] \\
     = &\; \y+h^2\M^{-1}\fin(\x_{n+1}),
\end{aligned}
\end{equation}
which is essentially the first-order optimality condition in Eqn. (\ref{eqn:implicit_energy}). Differentiating $\x_{n}$ on both sides gives
\begin{equation}\label{eqn:backprop_firststep}
    \begin{aligned}
    \frac{\partial \x_{n+1}}{\partial \x_n}
    = &\; \Imat + h^2 \M^{-1} \frac{\partial \fin(\x_{n+1})  }{\partial \x_{n+1}} \frac{\partial  \x_{n+1} }{\partial  \x_{n} } \\
    =&\; \Imat - h^2 \M^{-1} \nabla^2 W(\x_{n+1})\frac{\partial  \x_{n+1} }{\partial  \x_{n} },
    \end{aligned}
\end{equation}
or equivalently,
\begin{equation}\label{eqn:backprop_matrix}
    \begin{aligned}
    \frac{\partial \x_{n+1}}{\partial \x_n}
    = &\; [\Imat+h^2\M^{-1}\nabla^2 W(\x_{n+1})]^{-1} \\
    = &\; [\M + h^2\nabla^2 W(\x_{n+1})]^{-1}\M.
    \end{aligned}
\end{equation}
In backpropagation, such a Jacobian matrix is coupled with the gradients of a loss function $L$ which are passed to the previous state (assuming the gradients below are both column vectors):
\begin{equation}\label{eqn:adjoint}
\begin{aligned}
    \frac{\partial L}{\partial \x_n} = &\; (\frac{\partial \x_{n+1}}{\partial \x_n})^\top\frac{\partial L}{\partial \x_{n+1}} \\
    = &\; \M \underbrace{[\M + h^2\nabla^2 W(\x_{n+1})]^{-1} \frac{\partial L}{\partial \x_{n+1}}}_{\z_{n+1}}.
\end{aligned}
\end{equation}
Here, we obtain the adjoint vector $\z_{n+1}$ by solving the linear system of equations with the matrix $\M + h^2\nabla^2 W(\x_{n+1})$ on the left-hand side, which avoids an explicit inversion of the large and sparse matrix. Backpropagating gradients of $L$ with respect to $\vel_{n}$ and $\vel_{n+1}$ can be derived similarly. In fact, it solves the same linear system but with a different vector $\frac{\partial L}{\partial \vel_{n+1}}$ on the right-hand side.

\paragraph{Differentiating with Projective Dynamics} With assumptions on the energy form $W$ in PD, \citet{Du21_DiffPD} show that we can speed up the computation in $\z_{n+1}$ by exploiting the special form of $\nabla^2 W$. The first- and second-order derivatives in Eqn. (\ref{eqn:pd_energy}) are given by the following equations~\cite{Du21_DiffPD,Liu17_Quasi}:
\begin{align}
    \nabla W_i(\x) = &\; w_i\A_i^\top[\A_i\x - \p_i^*(\x)], \\
    \nabla^2 W_i(\x) = &\; w_i\A_i^\top\A_i - w_i\A_i^\top\frac{\partial \p_i^*}{\partial \x},
\end{align}
where $\p_i^*(\x)=\arg\min_{\p_i\in\mathcal{M}_i}\widetilde{W}_i(\x,\p_i)$ is the projection of $\A_i\x$ onto $\mathcal{M}_i$ obtained in the local step. Interested readers can refer to the appendix of~\citet{Liu17_Quasi} for their derivation details. We plug them to $\z_{n+1}$ and rewrite the linear system as follows:
\begin{align}\label{eqn:adjoint_pd}
    (\Pmat - \underbrace{h^2\sum_i w_i \A_i^\top \frac{\partial \p_i^*}{\partial \x}\Bigr|_{\x_{n+1}}}_{\Delta\Pmat})\z_{n+1}=\frac{\partial L}{\partial \x_{n+1}}.
\end{align}
Noting that a prefactorization of $\Pmat$ exists in Projective Dynamics,~\citet{Du21_DiffPD} propose the following iterations to solve $\z_{n+1}$:
\begin{align}\label{eqn:pd_jacobi}
    \Pmat\z_{n+1}^{k+1}=\Delta \Pmat\z_{n+1}^k + \frac{\partial L}{\partial \x_{n+1}}.
\end{align}
Similar to the local-global steps in Projective Dynamics, a local step can evaluate $\Delta \Pmat$ across each $\nabla^2 W_i$ in parallel, and a global step, which reuses $\Pmat$, can solve $\z_{n+1}^{k+1}$ with backsubstitution only.~\citet{Du21_DiffPD} show that such a local-global solver is empirically faster than directly solving Eqn. (\ref{eqn:adjoint}), which resembles the source of efficiency in the original PD method.

\subsection{Gradients with Contact}\label{sec:method:contact}

While~\citet{Du21_DiffPD} have discussed extensively how to fuse backpropagation into the Projective Dynamics framework, its support of contact is limited to non-penetration conditions without a proper treatment on friction. On the other hand, other prior papers have provided solutions to deriving gradients with contact governed by Signorini-Coulomb law, but they focus on either small-scale problems (e.g., rigid bodies in~\citet{De18_End2End}) or sparse contact events on a static ground only~\cite{Geilinger20_ADD}. Here, we present our differentiable cloth simulator that both inherits the speedup from PD and handles gradients with complicated contact events like self-collisions, which are overlooked in many existing differentiable simulators but common in cloth simulation.

\paragraph{Differentiating implicit time integration with contact} Consider the $n$-th time step in simulation with contact which takes as input $(\x_n, \vel_n)$ and computes $(\x_{n+1}, \vel_{n+1}, \rforce_{n+1}, \uu_{n+1})$ that satisfy Eqn. (\ref{eqn:implicit_with_contact}). In particular, for each contact node $j$, the forward simulation identifies which one of the contact conditions in Eqn. (\ref{eqn:contact_all}) applies to $(\rforce_{n+1,j}, \uu_{n+1,j})$. As an example, assuming a contact node $j$ satisfies Eqn. (\ref{eqn:contact_3}), its constraints can be summarized as follows:
\begin{subequations}
\begin{align}
    \|\rforce_{n+1,j|T}\| - \mu \rforce_{n+1,j|N} = &\; 0, \label{eqn:contact_backward_1}\\
    \uu_{n+1,j|N} = &\; 0, \label{eqn:contact_backward_2}\\
    (\uu_{n+1,j|T})_x(\rforce_{n+1,j|T})_y - (\uu_{n+1,j|T})_y (\rforce_{n+1,j|T})_x = &\; 0, \label{eqn:contact_backward_3}\\
    \uu_{j|T}\cdot\rforce_{j|T} \leq &\; 0,
\end{align}
\end{subequations}
where $(\cdot)_x$ and $(\cdot)_y$ extract the $x$ and $y$ components of a two-dimensional vector, respectively. The other two cases of Eqn. (\ref{eqn:contact_all}) similarly enforce three equality constraints ($\rforce_{n+1,j}=0$ for taking off and $\uu_{n+1,j}=0$ for sticking) and a number of inequality constraints on $(\rforce_{n+1,j}, \uu_{n+1,j})$. If the inequality constraint is inactive, slightly perturbing the inputs to the simulator will keep $(\rforce_{n+1,j},\uu_{n+1,j})$ inside its interior. Therefore, we can remove the inactive inequality when deducing the gradients during backpropagation. When the inequality constraint is active, the gradients of the simulation are not well defined because it represents corner cases that can be categorized into more than one contact types. These corner cases introduce non-smoothness to the simulator, but it is worth mentioning that they do not create discontinuities, just like the standard rectifier (ReLU) activation function is still continuous despite the non-smoothness at its turning point.

\hlred{After removing the inequality constraint, we further define a nonlinear vector function $\Cmat^j_{n}(\rforce_{n+1,j},\uu_{n+1,j})$ for the left-hand side of the three equality constraints and compactly represent the constraints as $\Cmat^j_{n}(\rforce_{n+1,j},\uu_{n+1,j})=0$.
}
It is convenient that $\Cmat^j_{n}$ for all three cases in Eqn. (\ref{eqn:contact_all}) have three dimensional outputs. We can stack $\Cmat^j_{n}$ from all contact nodes $j\in\ContactSet_n$ into a nonlinear function $\Cmat_n(\cdot,\cdot):\R^{3|\ContactSet_n|}\times\R^{3|\ContactSet_n|}\rightarrow\R^{3|\ContactSet_n|}$ (note that we only evaluate the function at valid input pairs that lie in $\Con^j$), allowing us to restate Eqn. (\ref{eqn:implicit_with_contact}) as follows:
\begin{equation}\label{eqn:backprop_with_contact}
   \left\{\begin{array}{lr}
    \vel_{n+1} = \vel_{n} + h\mathbf{M}^{-1}[\fin(\x_{n} + h\vel_{n+1})+\fex +\J_n^\top \rforce_{n+1}], \\
    \Cmat_n(\rforce_{n+1},\J_{n}\vel_{n+1})=\mathbf{0}.
     \end{array}\right.
\end{equation}
\hlred{Note that we choose $\vel_{n+1}$ instead of $\x_{n+1}$ as our variable in order to be consistent with the forward PD simulation with contact described by~\citet{Ly20_PDdryFriction}.} Differentiating both sides of the first equation with respect to $\vel_n$ leads to the following result:
\begin{align}
    \frac{\partial \vel_{n+1}}{\partial \vel_n} = &\; \Imat + h\M^{-1}[-h\nabla^2 W(\x_{n} + h\vel_{n+1}) + \J_n^\top \frac{\partial \rforce_{n+1}}{\partial \vel_{n+1}}]\frac{\partial \vel_{n+1}}{\partial \vel_n}\label{eqn:adjoint_vv} \\
    \frac{\partial \vel_{n+1}}{\partial \vel_n} = &\; [\Imat + h^2\M^{-1} \nabla^2 W(\x_n + h\vel_{n+1}) -h\M^{-1}\J_n^\top\frac{\partial \rforce_{n+1}}{\partial \vel_{n+1}}]^{-1} \\
    = &\; [\M + h^2\nabla^2 W(\x_{n} + h\vel_{n+1}) - \underbrace{ h\J_n^\top\frac{\partial \rforce_{n+1}}{\partial\vel_{n+1}}}_{\Delta \Rmat^\top}]^{-1}\M.
\end{align}
Comparing it with Eqn. (\ref{eqn:backprop_matrix}), we see the matrix to be inverted now has an additional component dependent on $\frac{\partial \rforce_{n+1}}{\partial \vel_{n+1}}$, which we obtain from differentiating the constraint $\Cmat_n=\mathbf{0}$ in Eqn. (\ref{eqn:backprop_with_contact}):
\begin{align}
    \frac{\partial \Cmat_n}{\partial \rforce}\Bigr|_{\rforce_{n+1}}\frac{\partial \rforce_{n+1}}{\partial \vel_{n+1}} + \frac{\partial \Cmat_n}{\partial \uu}\Bigr|_{\J_n\vel_{n+1}}\J_n = \mathbf{0}, \\
     \frac{\partial \rforce_{n+1}}{\partial \vel_{n+1}} = -(\frac{\partial \Cmat_n}{\partial \rforce}\Bigr|_{\rforce_{n+1}})^{-1} \frac{\partial \Cmat_n}{\partial \uu}\Bigr|_{\J_n\vel_{n+1}}\J_n.
\end{align}
\hlred{We stress that computing $\frac{\partial \Cmat_n}{\partial \rforce}$ and $\frac{\partial \Cmat_n}{\partial \uu}$ is trivial because both partial derivatives are $3\times3$ block-diagonal matrices. Therefore, $\Delta \Rmat$ can be parallelized among all contact nodes.} Backpropagation through $\vel_{n+1}$ to $\vel_n$ can be implemented with the same adjoint method before, which we give in the equation below for completeness with the notation $\z_{n+1}$ overloaded:
\begin{align}\label{eqn:adjoint_v}
    \frac{\partial L}{\partial \vel_n} = \M \underbrace{[\M + h^2\nabla^2 W(\x + h \vel_{n+1}) - \Delta \Rmat]^{-1}\frac{\partial L}{\partial \vel_{n+1}}}_{\z_{n+1}}.
\end{align}

\paragraph{Speedup with Projective Dynamics} With additional information about $W$ from PD, we can rewrite the adjoint vector $\z_{n+1}$ in Eqn. (\ref{eqn:adjoint_v}) by comparing it with Eqn. (\ref{eqn:adjoint_pd}):
\begin{align}\label{eqn:ppr_adjoint}
    (\Pmat - \Delta \Pmat - \Delta \Rmat)\z_{n+1}=\frac{\partial L}{\partial \vel_{n+1}},
\end{align}
which naturally leads to the following iterative solver:
\begin{align}
    \Pmat \z_{n+1}^{k+1}=(\Delta\Pmat + \Delta \Rmat)\z_{n+1}^k + \frac{\partial L}{\partial \vel_{n+1}}.
\end{align}
Comparing it with Eqn. (\ref{eqn:pd_jacobi}), we see the role of $\Delta\Pmat$ is replaced with $\Delta \Pmat + \Delta \Rmat$. Since we have shown that computing $\Delta \Rmat$ can be massively parallelized among contact nodes, it is suitable for contact-rich cloth simulation and preserves the efficiency of the local-global solver.

\paragraph{Convergence} In theory, such an iterative solver is guaranteed to converge from any initial guesses of $\z_{n+1}$ when the spectral radius $\rho[\Pmat^{-1}(\Delta \Pmat + \Delta \Rmat)] < 1$. Empirically, we notice in our cloth simulation that divergence is uncommon, especially when high-precision back-propagation is not required (Sec.~\ref{sec:applications}). When the iterative solver fails to converge, we switch back to a direct sparse matrix solver to solve Eqn. (\ref{eqn:ppr_adjoint}).

\paragraph{Extensions} We deliberately skipped the full definition of $\J_n$ in all equations above for a clearer presentation of the main idea behind our differentiable cloth simulation. We now elaborate on the role of $\J_n$ and discuss its implications on more complex collisions. \hlred{When the contact surface is static but non-planar with spatially varying surface normals, $\J_n$ is dependent on the positions of the nodes where the contact events occur. In this case, we estimate the contact normal based on the positions where the contact events occur, which are given by any collision detection algorithms. Replacing $\J_n$ with $\J_n(\x_{n}, \vel_{n})$ requires very minimal extra work to the gradients in Sec.~\ref{sec:method:contact}, as it contributes to the gradients with respect to $\vel_n$ in a straightforward manner.} Another type of collisions we consider is self-collisions between nodes. In this case, contact occurs between pairs of nodes where the contact normals are defined by the relative position between two nodes in the pair. Therefore, we can still define $\J_n$ as a function of $\x_n$ with each row block now consisting of two blocks of nonzero elements corresponding to the two contact nodes in the pair. Similarly, the gradient derivation remains unchanged \hlred{except that the dependencies between $\J_n$ and $(\x_n, \vel_n)$ need to be added to Eqn. (\ref{eqn:adjoint_vv})}. As we inherit from~\citet{Ly20_PDdryFriction} the contact model on nodes only, we also inherit its limitation of not handling other types of self-collisions like edge-edge collisions or vertex-face collisions. We leave the derivation of gradients with such cases as future work.

\section{Evaluations}\label{sec:evaluations}

\hlred{This section evaluates and discusses numerical properties of the proposed differentiable cloth simulation method in Sec.~\ref{sec:method}. We first evaluate its gradients by analyzing their source of non-smoothness and studying their usefulness in high-dimensional optimization problems. Next, we compare the dry frictional model with the contact model in the state-of-the-art differentiable cloth simulation method~\cite{Liang19_DiffCloth} and discuss their differences. Finally, we analyze the numerical properties of our iterative solver in back-propagation and compare its performance with a direct linear solver.}

\subsection{\hlred{Continuity and Smoothness}}\label{sec:evaluations:continuity}
\hlred{A fundamental assumption in any differentiable simulators is that the underlying physics system is smooth so that gradients can be well-defined. Eqn. (\ref{eqn:implicit_with_contact}) contains three possible sources of discontinuities and non-smoothness in the order of their occurrences: determining the contact set $\ContactSet_n$, computing $\J_n$ that represents the local contact frames, and choosing between three branches from $\Con^j_n$ in Eqn. (\ref{eqn:contact_all}). Below we discuss the effects from each of them in detail.}

\paragraph{Continuity of branches in contact conditions} First, we empirically demonstrate through an experiment below that switching between the three branches in Eqn. (\ref{eqn:contact_all}) does not create discontinuity in Eqn. (\ref{eqn:implicit_with_contact}). In particular, consider the $n$-th time step and assume that $\Con_n^j$ is fixed and that $\J_n$ is a continuous function of $(\x_n, \vel_n)$, if we treat Eqn. (\ref{eqn:implicit_with_contact}) as a function that takes as input $(\x_n,\vel_n)$ and returns $(\x_{n+1},\vel_{n+1})$, we will validate that this function is continuous. In other words, perturbing $(\x_n, \vel_n)$ a little will not cause jumps in the resultant $(\x_{n+1},\vel_{n+1})$ even though the corresponding $\rforce_{n+1}$ may need to switch between branches during the perturbation. The intuition is that the three branches in Eqn. (\ref{eqn:contact_all}) together define a connect set $\Con_n^j$ in which $(\rforce_{n+1,j},\uu_{n+1,j})$ from one branch can smoothly transition to another.

\begin{figure*}[ht]
    \centering
        \includegraphics[width=\textwidth]{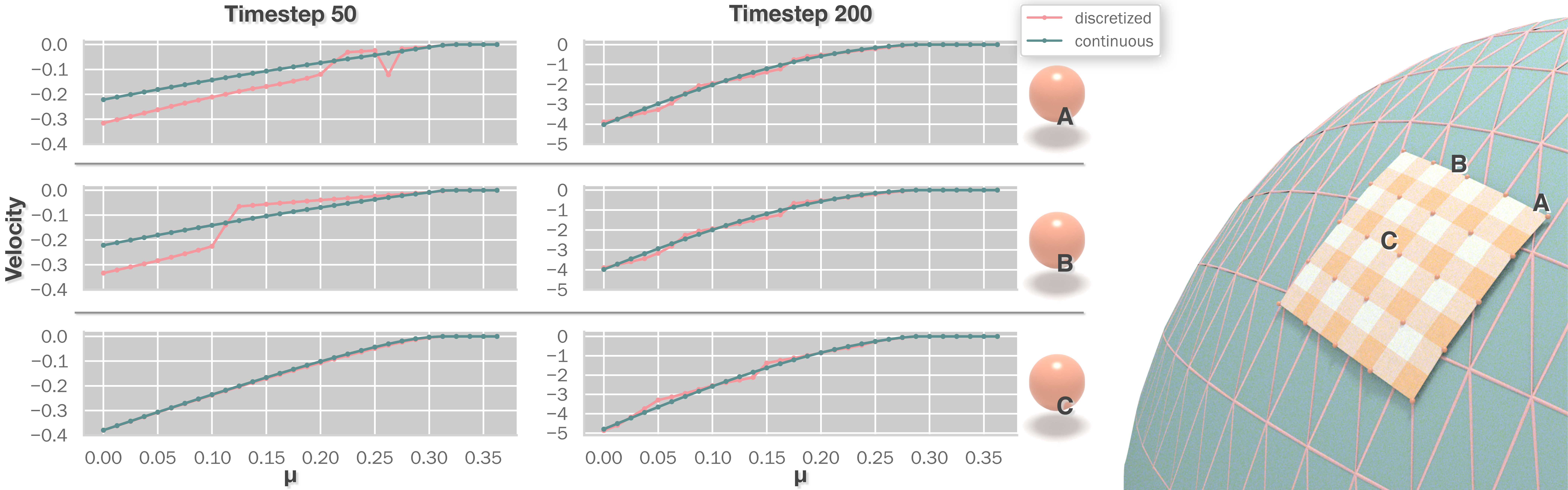}

    \caption{We simulate a piece of cloth sliding on a rigid sphere (right) to study the discontinuities and non-smoothness from contact conditions and surface geometry (Sec.~\ref{sec:evaluations:continuity}). The two columns of plots show the velocities of three contact nodes at the $50$-th and $200$-th time steps. The ``discretized'' and ``continuous'' labels indicate whether these velocities are computed with a triangulated sphere (visualized as pink triangles on the spherical surface) or a smooth sphere (visualized as green sphere surface).}
    \label{fig:continuousNormal}
\end{figure*}

We empirically validate the continuity of branch switching with the following experiment. We simulate a piece of cloth on a rigid, static, and frictional sphere for 200 timesteps. The sphere has one frictional coefficient $\mu$, and we repeat the experiments by varying $\mu$ from 0 to 0.35 (Fig.~\ref{fig:continuousNormal}). When $\mu$ is large, all nodes are fixed on the sphere due to their large static friction. When $\mu$ is close to $0$, the cloth slides on the sphere under gravity and takes off from the sphere near the end of the simulation. As $\mu$ changes from $0.35$ to $0$, each node on the cloth undergoes the transition from sticking to slipping, and eventually takes off. However, since each node has a different contact normal, the turning point for each of them to switch between these branches is different. Overall, when we gradually change $\mu$, the ratio among the numbers of nodes with static friction, dynamic friction, and no friction at the end of the simulation also changes gradually, allowing us to observe how switching between these branches affects the continuity of the physical quantities in simulation.

We summarize the quantitative results from this simulation experiment in Fig.~\ref{fig:continuousNormal} (green curves). Specifically, we plot the velocity of three nodes $A$, $B$, and $C$ as a function of $\mu$ at an intermediate time step (50) and near the end of simulation (200). We select three nodes from the corners, edge centers, and the center of the cloth, respectively. All velocities converge to $0$ when $\mu$ becomes large, which is as expected because a large $\mu$ leads to static friction that freezes these nodes. We also notice a turning point in the velocity curve for each node, indicating a switch between static and dynamic friction. The turning points are located at slightly different $\mu$ values for each node because their normals on the sphere surface are different. We observe from the figure that the branches in Eqn. (\ref{eqn:contact_all}) do not cause discontinuities.

While the above experiment confirms that these branches do not create discontinuities, we note that branch switching does introduce non-smoothness due to corner cases that can be arbitrarily classified into either branch, e.g., when a node is static but about to slip. Gradients at these corner cases are not well-defined, but the subset these corner cases occupy in $\Con^j_n$ has measure zero. Therefore, we still expect gradient-based optimization to be functional, just like we have observed the success of gradient-based optimization in modern deep learning with non-smooth but continuous operators, e.g., ReLU, max pooling, and so on.

\paragraph{Continuity of local frames at contact nodes} A second source of possible discontinuity and non-smoothness comes from computing $\J_n$, which consists of two steps: determining a contact point on the contact surfaces and calculating the local normal and tangent vectors. Both steps depend on the geometric representation of the contact surfaces. As pointed out by previous papers in rigid body dynamics~\cite{Popovic00_Manipulation,Werling-RSS-21}, contact surface discretization causes jumps in surface normals, and therefore they create discontinuities in velocity and position calculation. To confirm this observation in cloth simulation, we repeat the previous experiment by replacing the analytically described sphere with a triangulated one (pink triangulated sphere surface in Fig.~\ref{fig:continuousNormal} right). After such replacement, we clearly observe jumps in the intermediate and final velocity from the selected contact nodes (pink curves in Fig.~\ref{fig:continuousNormal}). Note that the jumps in velocities from the final velocities are less evident than from the intermediate velocities because the vertical axes have different ranges. This observation agrees with similar experiments from previous papers about rigid body dynamics and suggests that one should favor analytical surfaces in differentiable cloth simulation whenever possible.

\hlred{We end our discussion on the discontinuities from $\J_n$ with two more remarks. First, we notice the contact node velocity curves in Fig.~\ref{fig:continuousNormal} are partitioned into a small number of continuous segments, which is consistent with the result reported by previous work~\cite{Popovic00_Manipulation} discussing contact events on rigid body dynamics.
Second, if the scene contains multiple objects the cloth can be in contact with, it is not uncommon to see jumps in the locations of contact events, e.g., from one object to another, even with a continuous representation of each object. Such jumps naturally lead to large discontinuities in simulation. While we do not provide solutions to it in this work, a closely related issue has been extensively studied in differentiable rendering~\cite{Li2018DMC} from which we may borrow inspirations in the future.}

\hlred{\paragraph{Continuity of contact sets} The last and most common source of discontinuity comes from deciding the contact set at each time step, i.e., whether a node should be added to the contact set or not. Obviously, this selection process is not continuous. It is worth noting that having a constant contact set does not cause too much trouble for gradient computation regardless of its size (Fig.~\ref{fig:continuousNormal}). Instead, it is the \emph{change} in the contact set from time to time that brings in discontinuities because whenever a new node is put in the contact set, it adds an impulsive force to the right-hand side of Eqn. (\ref{eqn:implicit_with_contact}).}

\begin{figure}[htb]
    \centering
    \includegraphics[width=\columnwidth]{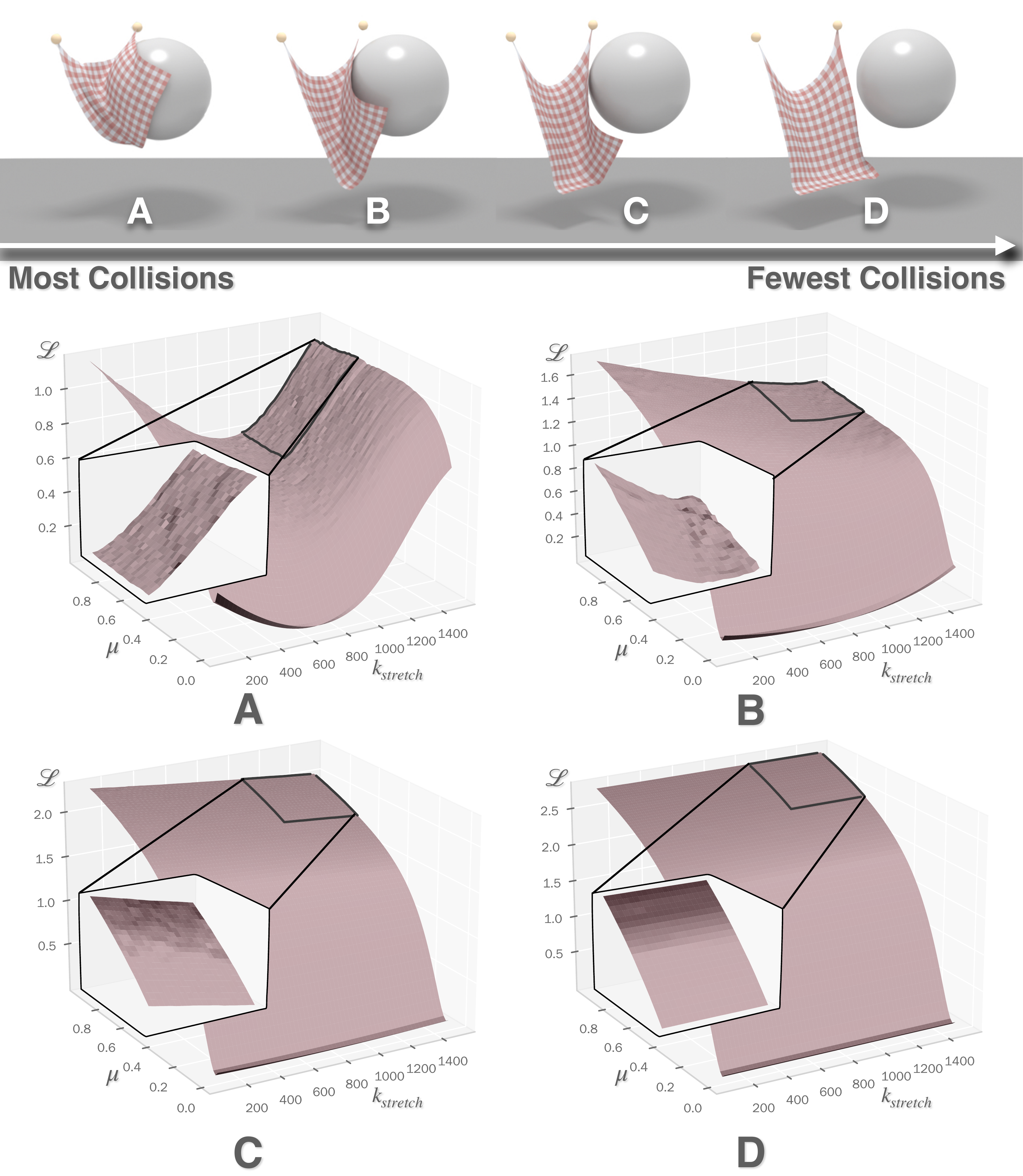}
    \vspace{-2em}
    \caption{\hlred{We simulate collisions between a piece of cloth and a rigid sphere at four different horizontal locations to study the effects of varying numbers of collisions on the smoothness of the simulation. Top: rendering of one of the time steps when the cloth is expected to be in contact with the sphere. The four scenes are ordered by their number of collisions. Bottom: landscapes of the loss defined as a function of the frictional coefficient $\mu$ and the stiffness $k$ of the cloth (Sec.~\ref{sec:evaluations:continuity}). The zoomed-in views of the landscapes show increasing discontinuity and non-smoothness as more collisions occur.}}
    \label{fig:evaluation_collision_num}
\end{figure}

\hlred{To better understand the effects of changes in contact sets, we hang a piece of cloth above a static sphere in simulation and let the lower half of the cloth fall and slide on the sphere due to gravity (Fig.~\ref{fig:evaluation_collision_num} top). We equip this experiment with a system identification task of the frictional coefficient $\mu$ and the stiffness parameter $k$ of the cloth: given a motion sequence of the cloth generated from a pair of unknown $\mu$ and $k$, we define a loss function that sums over all time steps the squared distance between each node position in simulation and its corresponding location in the given motion sequence. We repeat the task with four settings of the sphere at different horizontal offsets, leading to a varying frequency of contact events among them.}

\hlred{We plot the landscape of the loss function in Fig.~\ref{fig:evaluation_collision_num} (middle) and compare its smoothness among the four settings with different frequencies of contact events. At first glance, it seems that all four landscapes are equally smooth. However, magnifying a small region of each landscape shows that there is a profound distinction between their continuity and smoothness (Fig.~\ref{fig:evaluation_collision_num} bottom): as establishing and breaking contact becomes more frequent, the local landscape tends to be bumpier.}

\hlred{\paragraph{Summary} We have discussed the three sources of potential discontinuities and non-smoothness in our differentiable cloth simulator ordered by their damage to the gradients: the branches in the contact conditions only introduce non-smoothness and maintain continuity; contact surface discretization creates discontinuities due to the jumps of normals across adjacent triangles, but we still expect continuity almost everywhere; adding or deleting nodes in the contact set creates frequent and the most severe discontinuities due to the introduction or removal of impulsive forces.}

\subsection{\hlred{Usefulness of Gradients}}~\label{sec:evaluations:usefulness}

\begin{figure}[htb]
    \centering
    \includegraphics[width=\columnwidth]{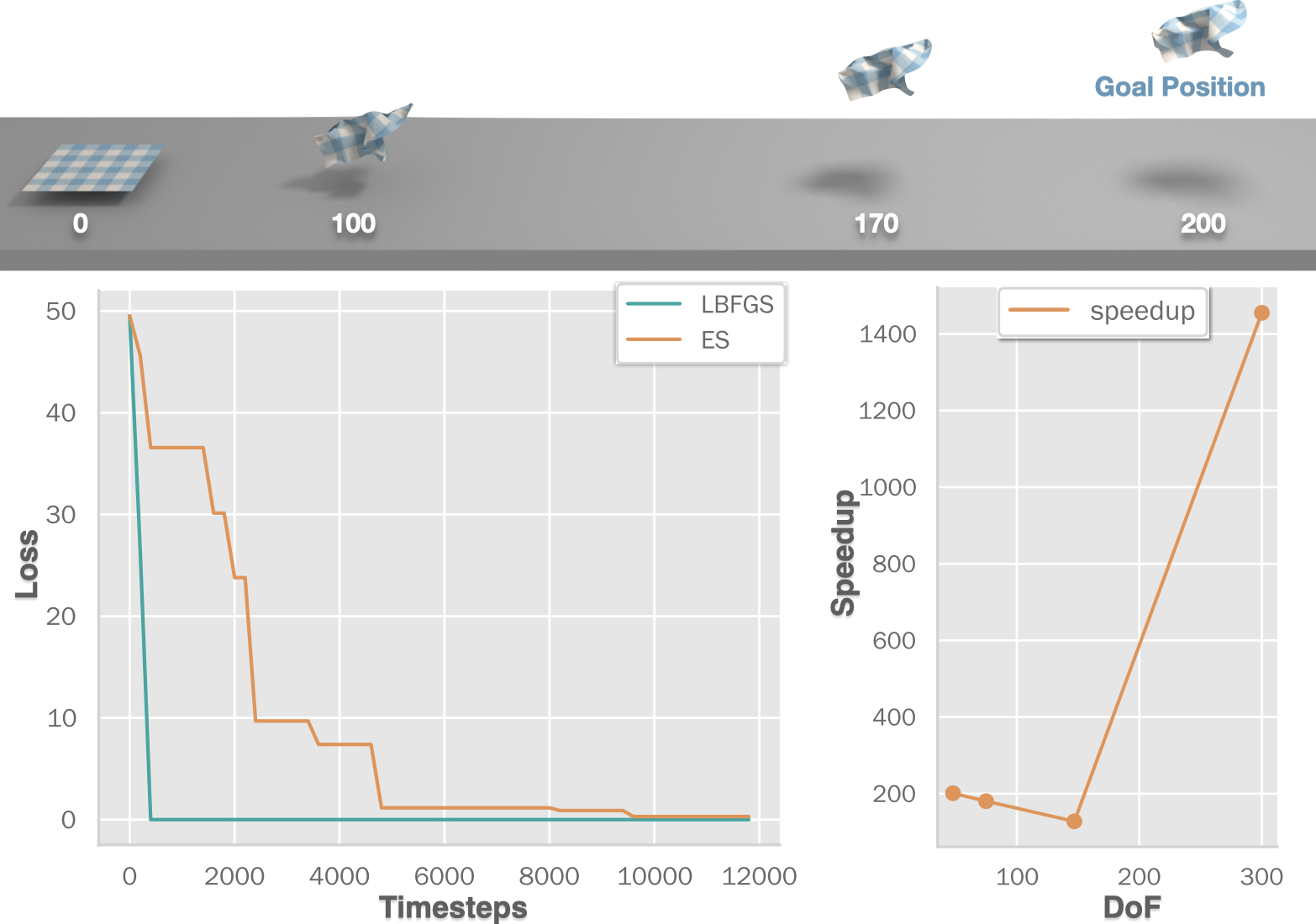}
    \vspace{-2em}
    \caption{\textbf{Flying Napkin}. \hlred{Comparisons between gradient-based and gradient-free methods for optimizing high-dimensional decision variables in differentiable cloth simulation. Top: the motion sequence of a piece of cloth with external forces parametrized with 300 variables. The goal is to find proper external forces with which the cloth ends at a target position. The loss function is defined as the position discrepancy between the simulated motion sequence and the reference motion. Bottom left: the loss vs. time step curves (300 DoF) from gradient-based (L-BFGS-B) and gradient-free (CMA-ES) methods. Both L-BFGS-B and CMA-ES use the same random seeds. Bottom right: we vary the degrees of freedom in the external force parametrization and repeat the experiment three times. For each experiment, we compute the ratio between the number of time steps used by gradient-free and gradient-based methods until they converge or use up the time budget and plot them as a speedup vs. DoFs curve.}}
    \label{fig:high_dim_opt}
\end{figure}

\hlred{One advantage of gradient-based approaches over their gradient-free counterparts is their faster convergence rate: typically, gradient-free methods explore the local landscape of an objective function by evaluating it with massive samples in the neighborhood. However, such a sampling approach quickly becomes less efficient as the dimension of the decision variables grows higher. Due to this reason, we hypothesize that using gradients from our differentiable simulator will become most beneficial when we have a large number of decision variables. We verify this hypothesis using a control optimization problem: we simulate a piece of cloth with time-invariant external forces applied to each node. The goal is to design the force at each node to pull the center of the cloth to a target position in the end (Fig.~\ref{fig:high_dim_opt} top). We simulate the cloth with four settings on its DoFs: $4\times4$ nodes, $5\times5$ nodes, $7\times7$ nodes, and $10\times10$ nodes. Therefore, the number of variables in the optimization problem is $48$, $75$, $147$, and $300$, respectively.}

\hlred{Fig.~\ref{fig:high_dim_opt} reports in this problem the convergence rates of L-BFGS-B~\cite{Liu89LBFGS} and CMA-ES~\cite{Hansen06cmaes}, two representative gradient-based and gradient-free approaches, with varying DoFs. We start with the largest setting of the cloth (300 DoFs) and run both L-BFGS-B and CMA-ES with five randomly chosen initial guesses on the force values in parallel. We then plot the loss vs. time step curves for both methods in Fig.~\ref{fig:high_dim_opt} bottom left. Here, the loss is the minimum loss from each method's five parallel runs as a function of the time steps they have consumed during the optimization process. It is clear to see the obvious speedup of L-BFGS-B over CMA-ES in this 300-DoF optimization problem, just as we expected.}

\hlred{Additionally, we repeat the experiment with three other settings of the cloth and plot the speedup vs. DoF curves in Fig.~\ref{fig:high_dim_opt} bottom right, where the speedup is the ratio between the number of time steps used by CMA-ES and L-BFGS-B when their losses reach 0.01. Again, we observe significant speedup from L-BFGS-B across the board, with the largest speedup from the cloth with the highest DoFs. This observation confirms the benefits of using gradients from our differentiable cloth simulator in inverse design problems.}

\subsection{Benefits of Dry Frictional Contact} 
\begin{figure}[htb]
    \centering
    \includegraphics[width=\columnwidth]{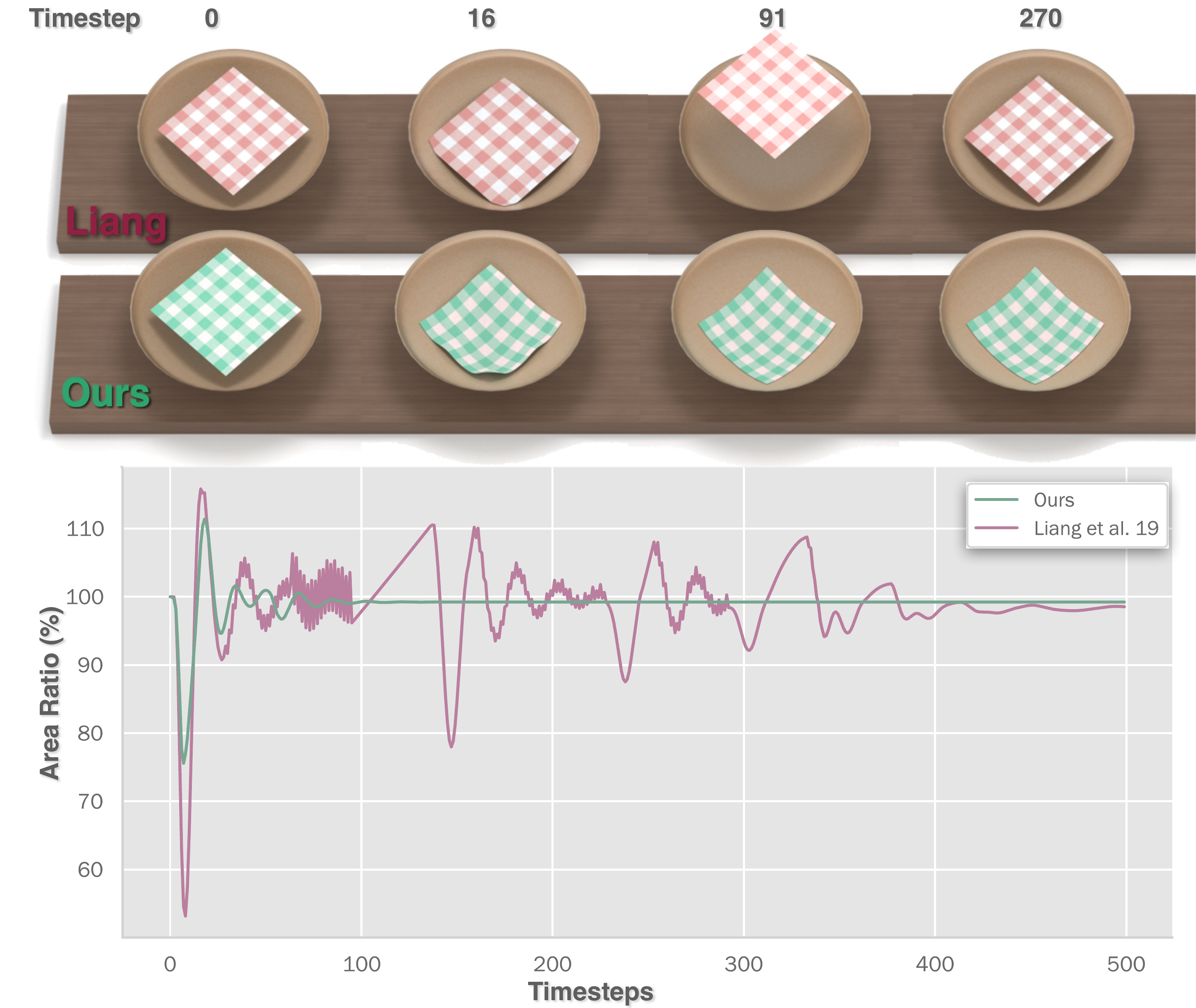} 
    \vspace{-2em}
    \caption{\textbf{Bowl}. We simulate the motion sequence of the napkin ($50\times50\times2$ triangles, $h=5$ms, $500$ time steps) with the contact models from two differentiable cloth simulators:~\citet{Liang19_DiffCloth} (top) and ours (middle). The area ratio (bottom) is the ratio between the napkin's area at the current timestep and in the rest shape. Refer to our video for the full motions.}
    \label{fig:evaluation_contact}
\end{figure}

We compare the contact model in our differentiable simulator with that in the state-of-the-art differentiable cloth simulator from~\citet{Liang19_DiffCloth}. Both models ensure penetration-free simulation and detect self-collisions (node-node collisions in ours and vertex-face and edge-edge collisions in~\citet{Liang19_DiffCloth}). However, the contact model in~\citet{Liang19_DiffCloth} does not take into account complementarity conditions on either contact forces or frictional forces, which may lead to undesirable artifacts. To visualize this issue, we consider a test scenario in which a napkin falls freely onto the inner surface of a bowl (Fig.~\ref{fig:evaluation_contact}). The differentiable simulators from both~\citet{Liang19_DiffCloth} and our work manage to simulate the napkin without numerical explosion. However, the dry frictional contact model in our differentiable simulator shows more physically realistic motion (Fig.~\ref{fig:evaluation_contact} middle), whereas the contact model from~\citet{Liang19_DiffCloth} leads to more drastic changes in the size of the napkin with a popping artifact after its contact with the bowl (Fig.~\ref{fig:evaluation_contact} top and bottom). These artifacts are explainable because the collision handling algorithm in~\citet{Liang19_DiffCloth} modifies node positions after penetration without verifying whether such an update requires sticky contact forces. When the napkin becomes in contact with the concave inner surface of the bowl, such a direct modification injects extra elastic energies. This experiment confirms supporting a more advanced contact and friction model in differentiable cloth simulation is both viable and beneficial.

\subsection{Evaluation of Iterative Solver in Backpropagation} 
\begin{figure}[htb]
    \centering
    \includegraphics[width=\columnwidth]{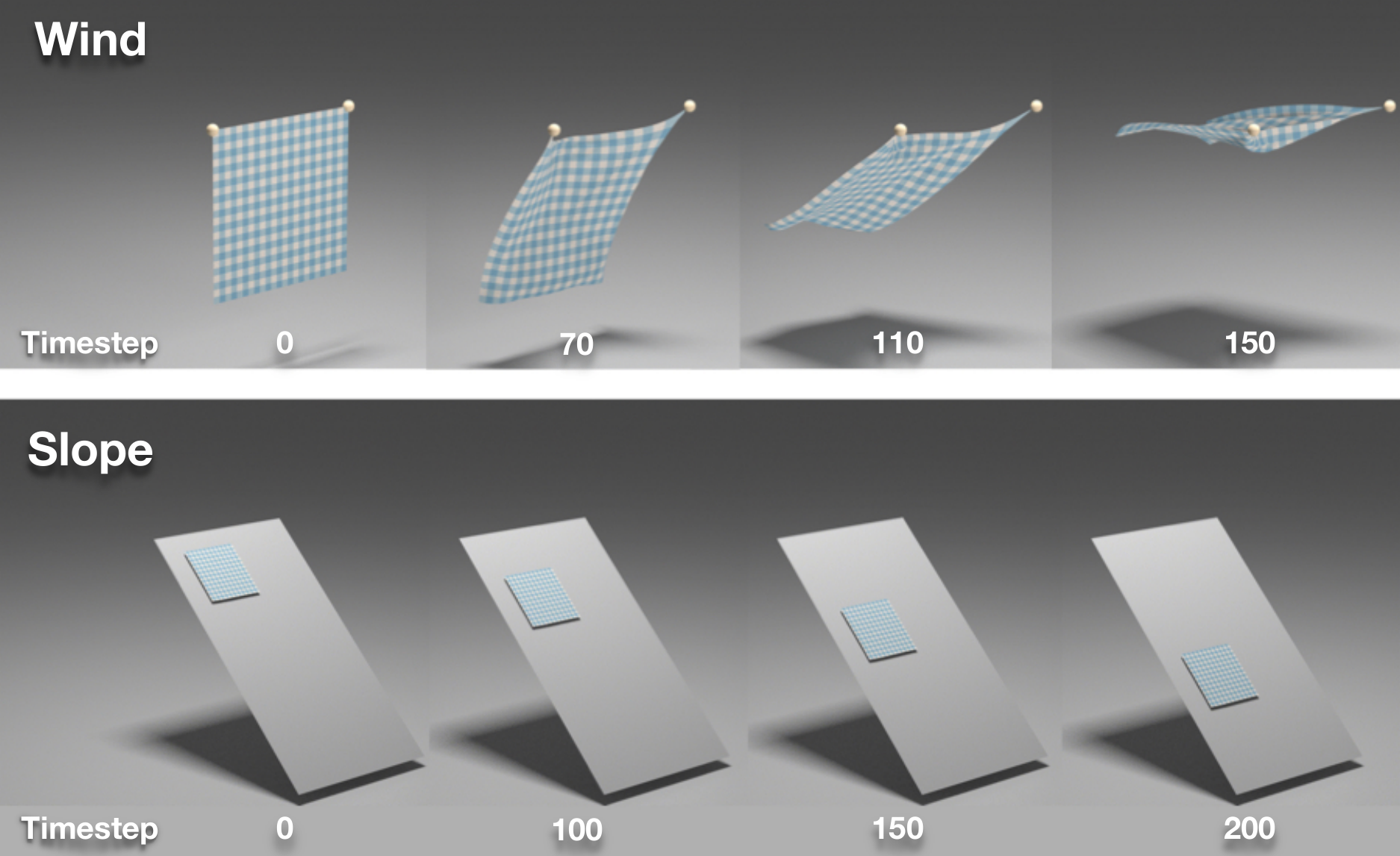}
    \caption{\textbf{Wind} and \textbf{Slope}. Motion sequences from the two benchmark tests for comparing iterative and direct solvers in back-propagation. Top: the ``Wind'' test ($h=1/90$s, 200 time steps) where a piece of cloth moves under synthetic wind. Bottom: the ``Slope'' test ($h=1/100$s, 300 time steps) where a ribbon slides along a slanted plane.}
    \label{fig:wind_and_slope}
\end{figure}

Another key component in our differentiable cloth simulation is the iterative solver in Eqn. (\ref{eqn:ppr_adjoint}) that utilizes the prefactorized $\Pmat$ in PD. In a standard differentiable simulator, solving Eqn. (\ref{eqn:ppr_adjoint}) would be done with a sparse matrix solver treating $(\Pmat - \Delta \Pmat - \Delta \Rmat)$ as a whole. To understand the performance of the proposed iterative solver, we design two benchmark tests: a ``Wind'' test where a hanging napkin moves under synthetic wind and a ``Slope'' test where a ribbon slides on a slanted plane (Fig.~\ref{fig:wind_and_slope}). These two tests represent two extreme cases in contact handling: the napkin in ``Wind'' barely has any contact or self-collisions, whereas every single node of the ribbon in ``Slope'' is in contact with the plane. We then compare the time cost of our iterative solver with a sparse LU solver in both tests. We implement both solvers using Eigen~\cite{Eigen18} and choose LU factorization because $(\Pmat - \Delta \Pmat - \Delta \Rmat)$ is usually not a symmetric or positive definite matrix, preventing us from using more specialized sparse matrix solvers like Cholesky factorization or Conjugate Gradient methods. For the iterative solver, we report two results from low-precision (1e-4) and high-precision (1e-6) convergence thresholds that control the termination condition in the iterations. We find these thresholds by varying it from 1e-1 to 1e-9 until the gradients computed from the iterative solver start to agree with the direct solver, which we treat as the ground-truth gradients. We repeat the experiments with three mesh resolutions to test the scalability of our method.

We summarize the statistics from both tests in Table~\ref{tab:windAndSlope}. Overall, the iterative solver in backpropagation is faster than the direct solver, and the speedup becomes more substantial as we increase the mesh resolution. The speedup is also more evident with low-precision threshold, which is consistent with the results reported in previous PD papers~\cite{Bouaziz14_PD,Liu17_Quasi,Du21_DiffPD}. A further decomposition of time confirms that solving the linear system takes up the majority of backpropagation time, so any improvement in the choice of solver has a dominating positive effect. Although the low-precision results may not match the ground-truth gradients as closely as their high-precision counterparts, their backpropagation is significantly faster and can still benefit gradient-based optimization. This is because even an imperfect gradient can still guide gradient descent algorithms to converge as long as it is along the descending direction. This is confirmed empirically by our experiments in Sec.~\ref{sec:applications}, in most of which we use low-precision convergence threshold and still observe successful gradient-based optimization results. It is also worth mentioning that contact-related operators, whose time cost is included in the ``Other'' and ``$\Delta\Rmat$'' columns in Table~\ref{tab:windAndSlope}, add very little extra overhead to the backpropagation. Finally, we observe uncommon but non-negligible failure cases of the iterative solvers in one experiment ($48\times48$ with 1e-6 as the threshold in Table~\ref{tab:windAndSlope}) ``Wind''. For the $15\%$ failed timesteps, we switch to the sparse LU solver, adding an extra $11.5\%$ time in backpropagation.

\begin{table*}[ht!]
\caption{ Comparison between the iterative solver (ours) and the sparse LU direct solver in backpropagation under various mesh resolutions in the ``Wind'' and ``Slope'' tests. The numbers in the ``Res.'' column report the mesh resolution. The number in parentheses in the ``Solver'' column indicates the epsilon value controlling the convergence of the Jacobi solver. The ``Backprop Time'' column reports the net time in backpropagation and is further decomposed into the next five columns, from which the sum of ratios is $100\%$: ``$\Delta \Pmat$'' and ``$\Delta \Rmat$'' report the time spent on assembling $\Delta \Pmat$ and $\Delta \Rmat$ respectively, ``Iter.'' and ``Direct'' shows the time cost by either solver, and operations not covered by these four columns are in the ``Other'' column. The ``Fail'' column reports the ratio between the number of timesteps seeing nonconvergence in our iterative solver and the number of total timesteps. The ``Speedup'' column is the ratio between the direct solver's time and our time in ``Backprop. Time'' in each test.}
\label{tab:windAndSlope}
\begin{tabular}{c|c|c|c|ccccc|c|c}
\toprule
\textbf{Test} & \textbf{Res.} & \textbf{Solver} & \textbf{Backprop Time (s)} & \textbf{$\Delta\Pmat$ (\%)} & \textbf{$\Delta\Rmat$ (\%)} & \textbf{Iter. (\%)} & \textbf{Direct (\%)} &  \textbf{Other (\%)} & \textbf{Fail (\%)} & \textbf{Speedup} \\
\midrule
\multirow{9}{*}{Wind} & \multirow{3}{*}{12x12} & Direct & 0.606 & 13.9 & \multirow{3}{*}{0} &  - & 84.4 & 1.7 & \multirow{3}{*}{0}  & - \\
& & Ours (1e-4) & 0.212 & 39.9 & & 55.5  & - & 4.6 &  & 2.9x \\
& & Ours (1e-6) & 0.424 & 19.2 & & 78.5 & - &  2.3 &  & 1.4x \\
 \cmidrule{2-11} 
& \multirow{3}{*}{24x24} & Direct & 3.360 & 9.9 & \multirow{3}{*}{0}  & - & 89.1  & 1.0 & \multirow{3}{*}{0} & - \\
& & Ours (1e-4) & 0.591 & 58.2 &  & 36.3 & -  & 5.5 &  & 5.7x \\
& & Ours (1e-6) & 2.858 & 11.7 &  & 87.2 & -  & 1.1 &  & 1.2x \\
 \cmidrule{2-11} 
& \multirow{3}{*}{48x48} & Direct & 24.001 & 6.8 & \multirow{3}{*}{0} & - & 92.5  & 0.7 & 0 & - \\
& & Ours (1e-4) & 2.262 & 63.2 & & 29.7 & - & 7.1 & 0 & 10.6x \\
& & Ours (1e-6) & 29.255 & 5.4 & & 82.6 & 11.5 & 0.6 & 15 & 0.8x \\
\midrule[0.1em]
\multirow{9}{*}{Slope} & \multirow{3}{*}{12x12} & Direct & 0.978 & 13.0 & 3.4 & - & 82.4 & 1.2 & \multirow{3}{*}{0}  & - \\
 & & Ours (1e-4) & 0.312 & 70.1 & 9.4 & 16.1  & - & 4.5 & & 3.1x \\
 & & Ours (1e-6) & 0.643 & 18.9 & 2.7 & 76.8 & - & 1.6  & & 1.5x \\
 \cmidrule{2-11} 
& \multirow{3}{*}{24x24} & Direct & 5.331 & 10.4 & 1.5  & - & 87.4 & 0.7 & \multirow{3}{*}{0}  & - \\
& & Ours (1e-4) & 0.781 & 70.1 & 9.4  & 16.1 & - & 4.5 & & 6.8x \\
& & Ours (1e-6) & 1.507 & 34.7 & 4.9 & 58.1 & -  & 2.4 & & 3.7x \\
 \cmidrule{2-11} 
& \multirow{2}{*}{48x48} & Direct & 37.296 & 6.5 & 0.7 & - & 92.5 & 0.4 & \multirow{3}{*}{0}  & - \\
& & Ours (1e-4) & 3.095 & 68.1 & 8.2 & 19.8 & - & 3.8 & & 12.0x \\
& & Ours (1e-6) & 6.825 & 31.2 & 3.5 & 63.6 & -  & 1.8 & & 5.5x \\
\bottomrule
\end{tabular}
\end{table*}

\section{Applications}\label{sec:applications}

In this section, we demonstrate a variety of cloth-related applications that can benefit from our proposed differentiable cloth simulation with dry frictional contact. \hlred{We repeat all experiments with multiple random seeds and use the same random seed set (therefore same initial parameter set) for all methods.  We report the optimization results and comparisons with other gradient-free algorithms in Table~\ref{tab:appLoss}. In Table~\ref{tab:appConv}, We report the time steps used by each optimization method to reach minimum final loss as well as the convergence ratio of our iterative solver during back-propagation. \hlrevision{For gradient-free algorithms, the number of time steps counts the total steps of forward simulation (each steps the simulation forward in time by $h$). For our gradient-based method, we double this number to include the number of back-propagation steps. Note that in practice, back-propagation takes much less time than forward simulation. We have included the wall clock time for running forward simulation and back-propagation of all examples in Appendix~\ref{sec:appendixExpRuntime}. In Table~\ref{tab:appLoss} and all loss vs. time steps plot below, we report the minimum loss or plot the minimum loss envelop achieved across all random seeds. See Appendix~\ref{sec:appendixAllSeedsExp} for the complete optimization results for each random seed.} Below we describe each application and highlight major results. } More information of each example is detailed in Appendix~\ref{sec:appendixExpDetails}. 
\paragraph{Implementations.} We write the backbone of our simulator in C++ and use Eigen for matrix and vector operations. Each application defines an optimization problem which we solve with L-BFGS-B, a classic gradient-based optimizer that can leverage the differentiability of our cloth simulator \hlred{while limiting parameters to physically-plausible ranges}. \hlred{We use the implementation of \textit{LBFGS++}~\cite{Qiu21_LBFGSpp} for L-BFGS-B, which implements the Mor\'{e}-Thuente line search~\cite{MoreThuente94_Linesearch}.} 

\begin{table*}
\caption{ \hlred{Comparison between the performance of gradient-based and gradient-free optimization on all examples (except ``Hat Controller'') in Sec.~\ref{sec:applications}. The ``Param \#'' and ``Seed \#'' columns report the number of optimization parameters and random seeds used in the experiments, respectively. All methods start the optimization with the same initial random seed set.  The ``Min Init. Loss'' column reports the minimum initial loss across all random seeds. The ``Optimized Loss Percentage'' column reports the optimized loss \emph{as a percentage} (0-100\%) of the minimum initial loss reached across all random seeds for each method: ``Final'' reports the loss reached when the optimization stops, and ``Equal \hlrevision{Step $\#$}'' reports the loss reached by CMA-ES and (1+1)-ES using the same number of time steps for L-BFGS-B convergence. We color the values of loss percentage using a green-orange color scale: green corresponds to 0\% and orange corresponds to 100\%.
}}
\label{tab:appLoss}
\begin{tabular}{c|c|c|c|c|c|c|c|c}
\toprule
   &   &   &   & \multicolumn{5}{c}{\textbf{Optimized Loss Percentage}}    \\ \cline{5-9} 
    &   &   &   & \textbf{L-BFGS-B (Ours)}  & \multicolumn{2}{c|}{\textbf{CMA-ES}}   & \multicolumn{2}{c}{\textbf{(1+1)-ES}} \\ \cline{5-9} 
 \multirow{-3}{*}{\textbf{Name}}  & \multirow{-3}{*}{\textbf{Param \#}} & \multirow{-3}{*}{\textbf{Seed \#}} & \multirow{-3}{*}{\textbf{Min Init. Loss}} & \textbf{Final (\%)}           & \textbf{Final (\%)}           & \textbf{Equal \hlrevision{Step $\#$} (\%)}      & \textbf{Final (\%)}           & \textbf{Equal \hlrevision{Step $\#$} (\%)}      \\  \midrule
T-shirt                         & 6                                   & 5                                      & 6.62                                       & \cellcolor[HTML]{83C4AE}0.53  & \cellcolor[HTML]{83C4AE}0.60  & \cellcolor[HTML]{E9E1AB}15.78 & \cellcolor[HTML]{8EC8AE}1.74  & \cellcolor[HTML]{95CAAE}2.43  \\
Sphere                          & 1                                   & 10                                     & 0.46                                       & \cellcolor[HTML]{82C4AE}0.43  & \cellcolor[HTML]{7EC3AE}0.00  & \cellcolor[HTML]{82C4AE}0.43  & \cellcolor[HTML]{84C4AE}0.65  & \cellcolor[HTML]{94C9AE}2.37  \\ 
Hat                             & 18                                  &         5                               & 21.83                                      & \cellcolor[HTML]{82C4AE}0.42  & \cellcolor[HTML]{83C4AE}0.55  & \cellcolor[HTML]{EAD69E}37.51 & \cellcolor[HTML]{82C4AE}0.51  & \cellcolor[HTML]{B6D3AE}5.87  \\ 
Sock                            & 36                                  & 5                                      & 17.08                                      & \cellcolor[HTML]{E9E3AD}11.59 & \cellcolor[HTML]{E9E0AA}17.39 & \cellcolor[HTML]{ECC98D}64.70 & \cellcolor[HTML]{E9E0A9}18.43 & \cellcolor[HTML]{EBCF94}52.80 \\ 
Dress                           & 2                                   & 16                                     & 0.90                                       & \cellcolor[HTML]{EDC284}76.91 & \cellcolor[HTML]{EDC284}79.02 & \cellcolor[HTML]{EDC184}80.02 & \cellcolor[HTML]{EEBA7B}77.25 & \cellcolor[HTML]{EDBB7C}82.57 \\ 
Flag                            & 8                                   & 10                                     & 2.88                                       & 
\cellcolor[HTML]{ACD0AE}4.77  & \cellcolor[HTML]{E4E1AE}10.57 & \cellcolor[HTML]{EAD69D}38.89 & \cellcolor[HTML]{E9DCA5}25.57 & \cellcolor[HTML]{EAD59C}40.14  \\  \bottomrule
\end{tabular}
\end{table*}

\begin{table}[htb]
\caption{\hlred{The ``Convergence Time Steps'' column reports the number of simulation time steps used for each optimization method until convergence on all examples shown in Sec.~\ref{sec:applications}. For a fair comparison, the time steps shown for L-BFGS-B include both forward simulation and backward propagation time. The ``Iter. Conv. \%'' column reports the percentage of time steps in backward propagation where the iterative solver converges. }}

\label{tab:appConv}
\begin{tabular}{c|c|c|c|c}
\toprule
\multirow{2}{*}{\textbf{Name}} & \multicolumn{3}{c|}{\textbf{Convergence Time Steps}} & \multirow{2}{*}{\textbf{Iter. Conv. \%}}        \\ \cline{2-4} 
                               & \textbf{L-BFGS-B} & \textbf{CMA-ES} & \textbf{(1+1)-ES}  &\\  \midrule
T-shirt  & 4500             & 53500           & 47750      &  99.75  \\ 
Sphere   & 2450             & 18900           & 11200      &  100.00      \\  
Hat      & 11200            & 159200          & 106000     &  90.58          \\  
Sock     & 17600            & 176800          & 91600      &  99.41          \\  
Dress    & 2750             & 3125            & 10000      &  84.80      \\  
Flag     & 6800             & 17300           & 24800      &  96.00          \\ \bottomrule
\end{tabular}
\end{table}

\subsection{System Identification}

\begin{figure}
    \centering
    \includegraphics[width=\columnwidth]{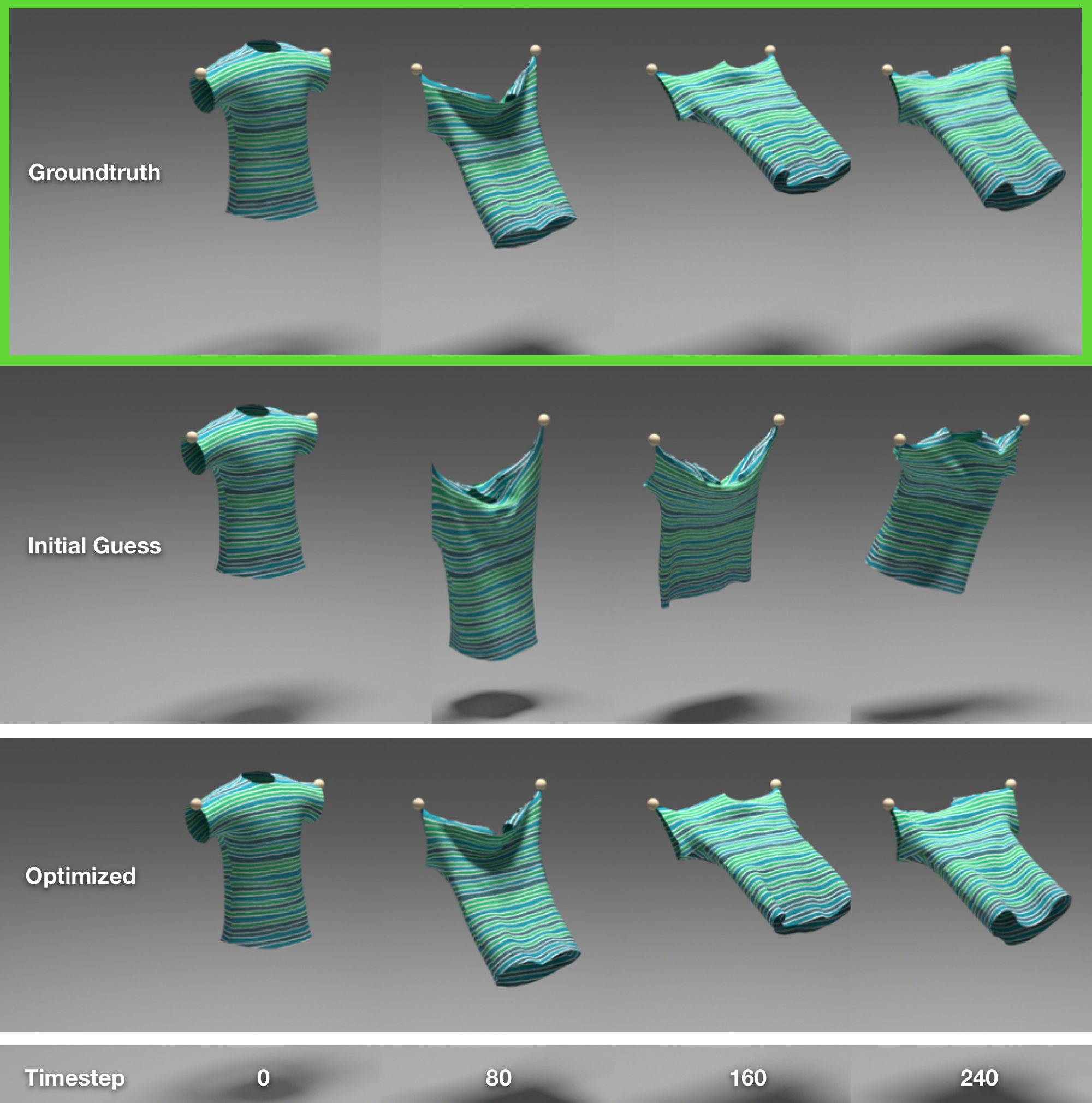}
    \vspace{-1.5em}
    \caption{\textbf{T-shirt} (4278 DoFs, $h=1/90$s, 250 time steps). Estimating the cloth material and wind parameters based on a given synthetic motion sequence of the T-shirt. From left to right we show the simulated T-shirt at 0, 80, 160, and 240 time steps. Top: the ground-truth motion sequence; Middle: simulation with the initial guess on the cloth and wind parameters; Bottom: simulation after optimization with L-BFGS-B.}
    \label{fig:tshirt}
\end{figure}

\begin{figure}
    \centering
    \includegraphics[width=\columnwidth]{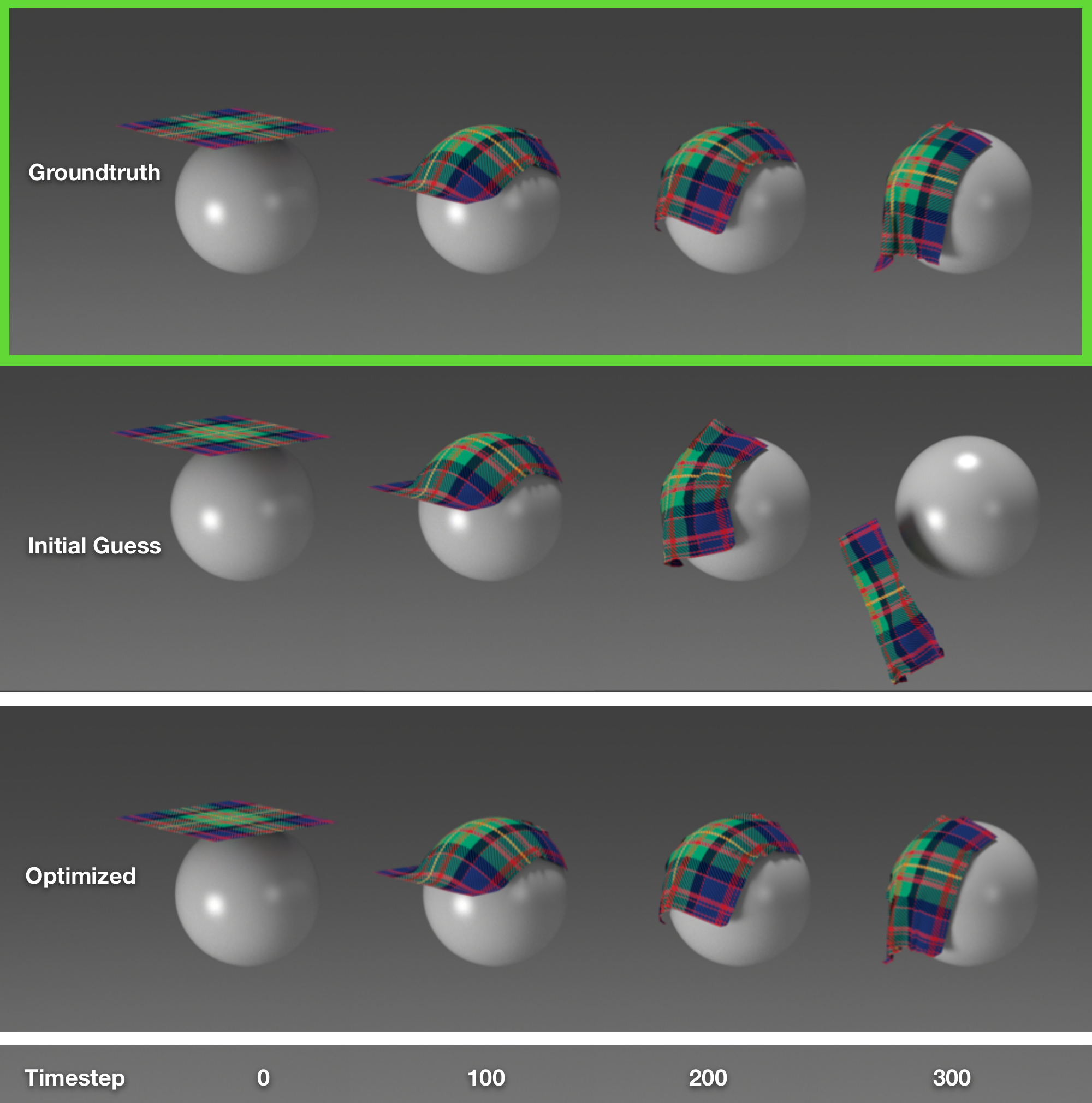}
    \caption{\textbf{Sphere} (1875 DoFs, $h=1/180$s, 350 time steps). Estimating the frictional coefficient $\mu$ \hlrevision{between the sphere and the cloth} based on the cloth's contact with the sphere. From top to bottom we show the simulation at 0, 100, 200, and 300 time steps. Top: the ground-truth motion sequence; Middle: simulation with the initial guess on $\mu$; Bottom: simulation after optimization with L-BFGS-B.}
    \label{fig:sphere}
\end{figure}

We start by showing two system identification examples: ``T-shirt'' (Fig.~\ref{fig:tshirt}) and ``Sphere'' (Fig.~\ref{fig:sphere}).

\paragraph{T-shirt}
In the ``T-shirt'' example, we are given a sequence of motions of a hanging T-shirt under synthetic wind generated from a parameterized sinusoidal function. The goal is to estimate a material parameter in the cloth (1 DoF) and identify the wind model parameters (5 DoFs controlling the amplitude, phase, and frequency of the sinusoidal function in three dimensions) from the motion data. 
We define the loss function as the L2-distance between the nodal positions of the T-shirt from the simulation and the given motion sequence. 

Unlike the ``Wind'' example in Sec.~\ref{sec:evaluations}, contacts and frictions are much more frequent in this example due to self-collision between the front and back layer of the T-shirt under the wind. We show the simulation of the T-shirt with parameters before and after optimization in Fig.~\ref{fig:tshirt} and compare the three optimizers in Table~\ref{tab:appLoss} and Table~\ref{tab:appConv}: L-BFGS-B, CMA-ES, and (1+1)-ES. Both CMA-ES and (1+1)-ES are standard gradient-free evolutionary strategies (ES).  We can conclude from Fig.~\ref{fig:tshirt} and Table~\ref{tab:appLoss} and ~\ref{tab:appConv} that all three methods manage to optimize system parameters leading to motion sequences visually identical to the given input, but L-BFGS-B converges much faster due to the extra knowledge of gradients. 

\paragraph{Sphere}
To highlight the effect of dry frictional contact in our simulator, we create the ``Sphere'' example with the goal of matching the motion sequence of a cloth interacting with a sphere by estimating the frictional coefficient \hlrevision{between the sphere and the cloth}. In this example, we let a piece of square cloth fall freely on a sphere, whose motion after being in contact with the sphere is largely controlled by the frictional coefficient. This example involves both self-collisions between nodes on the cloth and external contacts with the sphere. Similar to the example above, we run both L-BFGS-B and two ES baselines and report their statistics in Table~\ref{tab:appLoss} and ~\ref{tab:appConv}. All methods can optimize to a frictional coefficient that generates a motion sequence visually identical to the given input. However, this optimization problem is special in that it only has one parameter \hlred{and the landscape of the loss function is bumpy due to the large variation in the collision set as a function of the frictional coefficient, as suggested by the \textit{Continuity of contact sets} experiment in Sec.~\ref{sec:evaluations:continuity}}. In this case, evolutionary strategies can achieve a lower final loss because the samples can freely explore the full range of frictional contact, while L-BFGS-B can get trapped in a local optimum. 

\subsection{Robot-Assisted Dressing}\label{sec:robot-assited-dressing}

\begin{figure}[htb]
    \centering
    \includegraphics[width=\columnwidth]{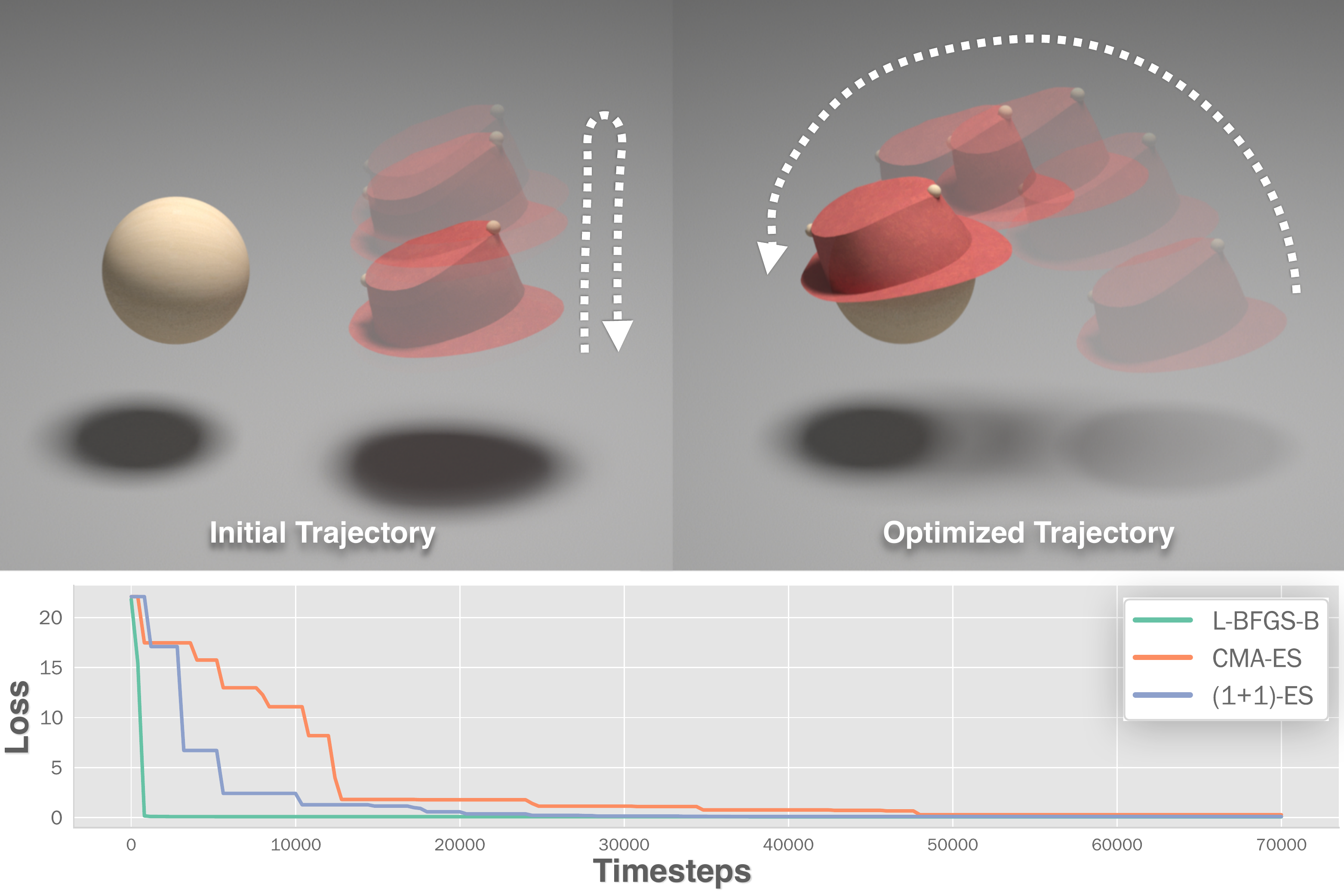}
    \caption{\textbf{Hat} (1737 DoFs, $h=1/100$s, 400 time steps). Optimizing trajectories for a manipulator to move a hat onto the sphere. Top left: Initial trajectory from one random seed before optimization overlaid with intermediate hat positions in simulation. Top right: the optimized trajectory from L-BFGS-B (which shares a visually similar trajectory with the ones optimized by ES algorithms). Bottom: The loss vs. time step curves for all methods.}%
    \label{fig:hat}
\end{figure}

\begin{figure}[htb]
    \centering
    \includegraphics[width=\columnwidth]{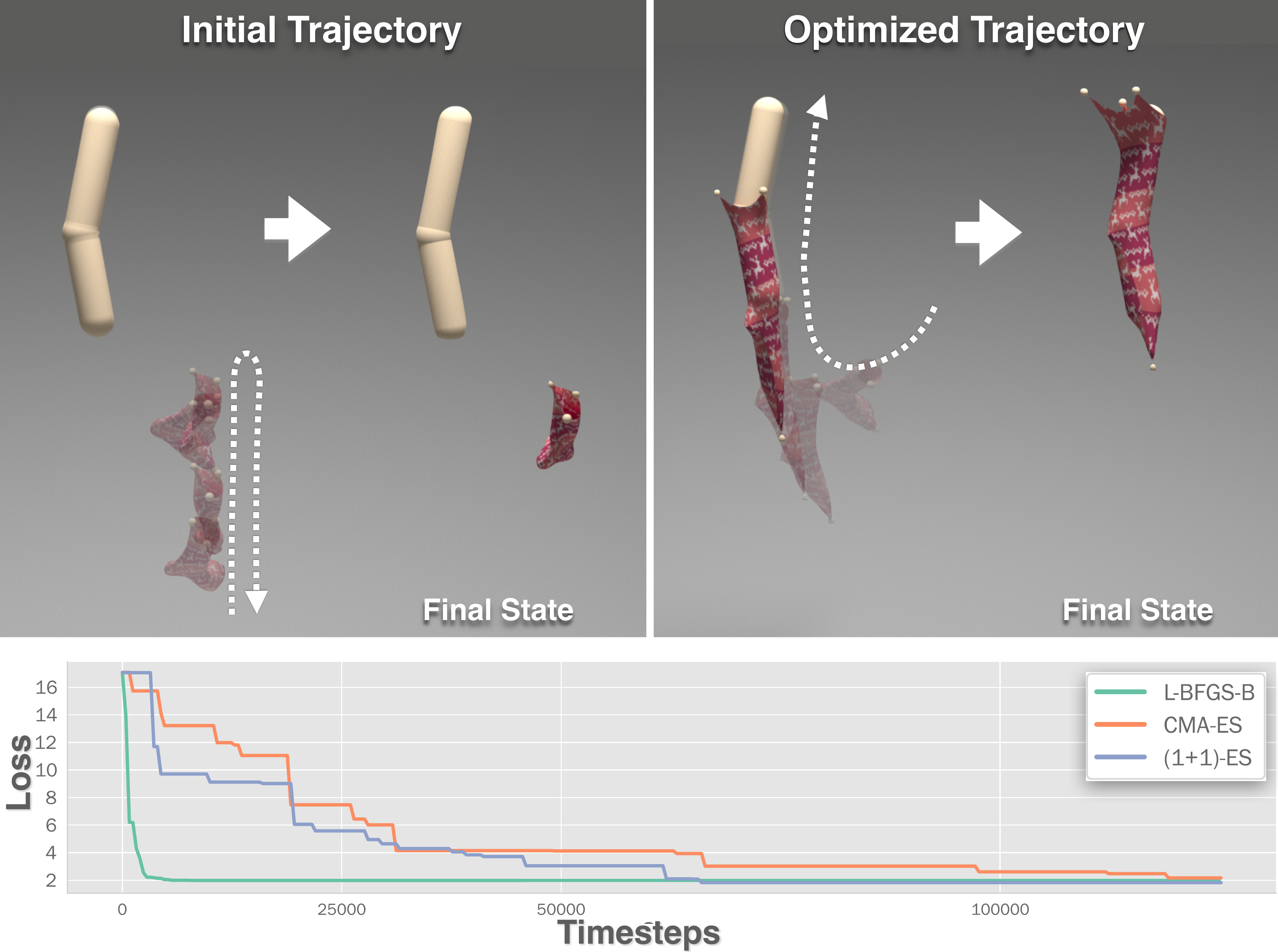}
    \caption{\textbf{Sock} (3165 DoFs, $h=1/160$s, 400 time steps). Optimizing trajectories for a manipulator to put a sock onto the foot model. Top left: One initial trajectory before optimization. We show the intermediate time steps on the left and the final state of the sock on the right. Top right: one control trajectory optimized by L-BFGS-B. The end effectors successfully put the sock onto the foot using the optimized trajectory. Bottom: the loss vs. time step curves for all methods.}%
    \label{fig:sock}
\end{figure}

Another line of research that can benefit from a differentiable cloth simulator is robot-assisted dressing. The mainstream solution to these tasks is typically gradient-free methods like evolutionary strategies, reinforcement learning, or inverse dynamics before a differentiable cloth simulator becomes available. We present two examples to demonstrate the usage of gradients in robot-assisted dressing: ``Hat'' (Fig.~\ref{fig:hat}) and ``Sock'' (Fig.~\ref{fig:sock}). In both examples, the goal is to find trajectories for a kinematic robotic manipulator to put on the hat or the sock. The end effectors of the manipulator pick a few prespecified vertices on the cloth meshes and pull them along the kinematic trajectories, which are parametrized as B-splines. By optimizing the parameters of the B-splines (18 DoFs in ``Hat'' and 36 DoFs in ``Sock''), we can direct the manipulator to move the hat until it reaches the target location on top of the sphere. We define the loss function in ``Hat'' as the L2-distance between the hat's final position at the last time step and a predefined target position, which we generate by translating the hat's rest shape onto the top of the sphere. The loss function for ``Sock'' is defined as the L2-distance between the desired and simulated locations of a few key points manually chosen on the sock and evaluated at the middle and the end of the simulation.

With the gradient information at hand, we run the L-BFGS-B optimizer to tune the parameters of the trajectories and compare its performance with CMA-ES and (1+1)-ES (Table~\ref{tab:appLoss} and~\ref{tab:appConv}). We notice that within the same \hlrevision{time steps}, L-BFGS-B converges substantially faster than the gradient-free baselines to a better solution (Fig.~\ref{fig:hat} and Fig.~\ref{fig:sock}). We can safely conclude that the fast convergence of L-BFGS-B unlocked by our differentiable cloth simulator is a clear advantage over gradient-free methods. 

\subsection{Inverse Design}

The next application we present is ``Dress'', an inverse design example that aims to optimize cloth material parameters in a dress so that its dynamic motion can satisfy certain design intents. Specifically, we optimize the material parameters of a twirl dress so that after the dress spins, the apex angle of the cone-like dress agrees with the target value (100 degrees in our experiment). We define the loss function as the difference between the hemline height corresponding to the target apex angle and the estimated apex angle from points on the hemline of the dress at the last frame of the simulation. We report the optimization results from L-BFGS-B and two ES baselines in Table~\ref{tab:appLoss} and visualize the simulation results before and after optimization in Fig.~\ref{fig:teaser}. Similar to the previous tasks, we notice that L-BFGS-B achieves better optimized results using fewer time steps.

\subsection{A Real-to-Sim Example}
\begin{figure}[htb]
    \centering
    \includegraphics[width=\columnwidth]{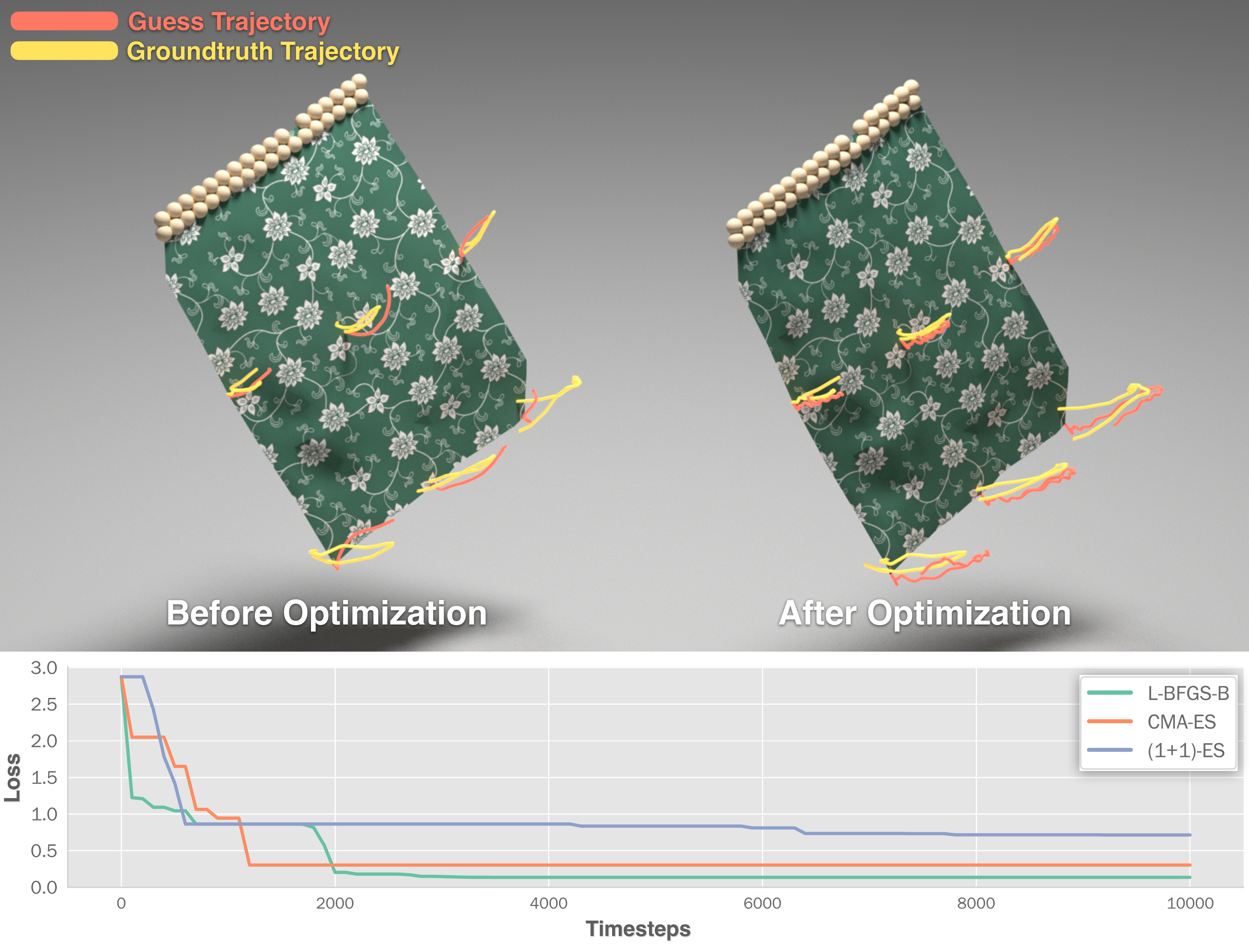}

    \caption{\textbf{Flag}. A real-to-sim example that reconstructs a digital flapping flag ($540$ vertices, $1026$ triangles, $h=1/120$s, 100 time steps) based on motion data captured from real-world experiments. Top: We plot the trajectories of 6 nodes on the cloth from the ground-truth motion (yellow curves) and the simulation results with guesses on the material and wind parameters before optimization (orange curves, left) and after optimization (orange curves, right). Bottom: The loss vs. time step curves for all methods.} %
    \label{fig:flag}
\end{figure}

In this section, we present a real-to-sim ``Flag'' example (Fig.~\ref{fig:flag}). In this example, we use the real-world motion sequence captured on a flag flapping in the wind from previous work~\cite{White07_Dataset} and aim to reconstruct a digital twin of the scene in simulation. This includes not only estimating the material parameters of the flag but also modeling the unknown wind condition at the capture time, which is particularly challenging due to its intricate stochastic model with unknown degrees of freedom. We model the wind force at each time step as a 3D force applied near the center of the scene and spatially decaying proportional to the inverse distance to the center. To model the transient nature of the wind force, we modulate magnitude of the 3D vector of a sinusoidal wave as a function of time with parameterized frequency and phase offset. Together, the material and wind model define an eight-dimensional parameter space to be optimized. We define the loss function as the L2-distance between the positions of all nodes at each time step in simulation and the ground-truth motion sequence.

We solve this task using L-BFGS-B and the two ES methods and report their performance in Table~\ref{tab:appLoss},~\ref{tab:appConv}, and Fig.~\ref{fig:flag}. All three methods substantially reduce the loss after optimization, but L-BFGS-B achieves a lower final loss. We plot the trajectories of 6 key nodes before and after optimization (orange) along with the ground-truth reference trajectories (yellow) from the motion-captured data in Fig.~\ref{fig:flag}. By comparing the left and right images in Fig.~\ref{fig:flag}, we can see that the L-BFGS-B optimizer reduces the discrepancy substantially between the simulated and actual trajectories after optimization. The real-to-sim matching is still imperfect as indicated by the nonzero final loss, which we suspect could be due to the simplistic nature of our synthetic wind model. A more sophisticated and expressive wind model, e.g., a neural network, may serve a better role in modeling and matching the real-world physics, which we leave as future work.

\begin{figure*}[htb]
    \centering
    \includegraphics[width=\textwidth]{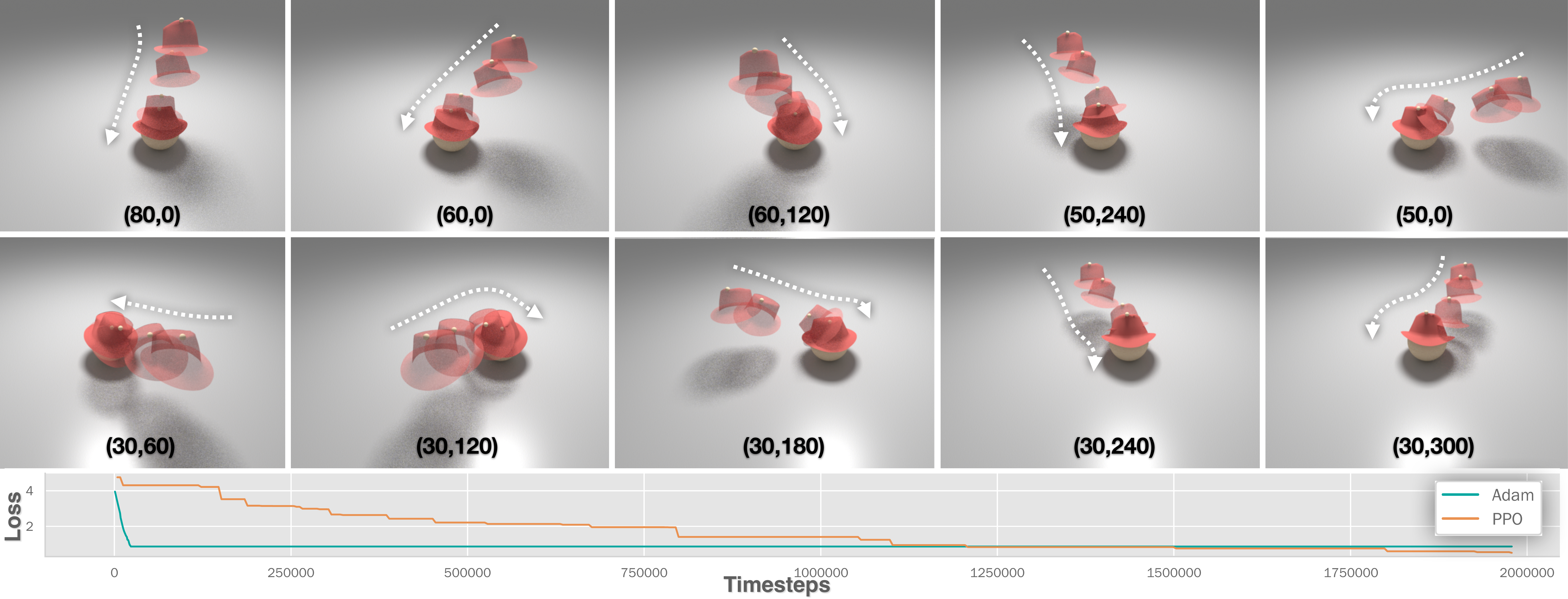}
    \caption{\hlred{\textbf{Hat Controller.} We train a closed-loop controller for the advanced ``Hat'' task (Sec.~\ref{sec:hat_controller}) which aims to move the hat from different initial positions (sampled from a fixed-radius hemisphere) onto the head. Top two rows: We visualize the control trajectories of the hat in ten initial positions denoted by their elevation and azimuth angle at the bottom of each subfigure. Bottom: The loss vs. time step curve for our gradient-based optimization and PPO.}}%
    \label{fig:hatController_sequences}
\end{figure*}

\begin{figure}[htb]
    \centering
      \includegraphics[width=\columnwidth]{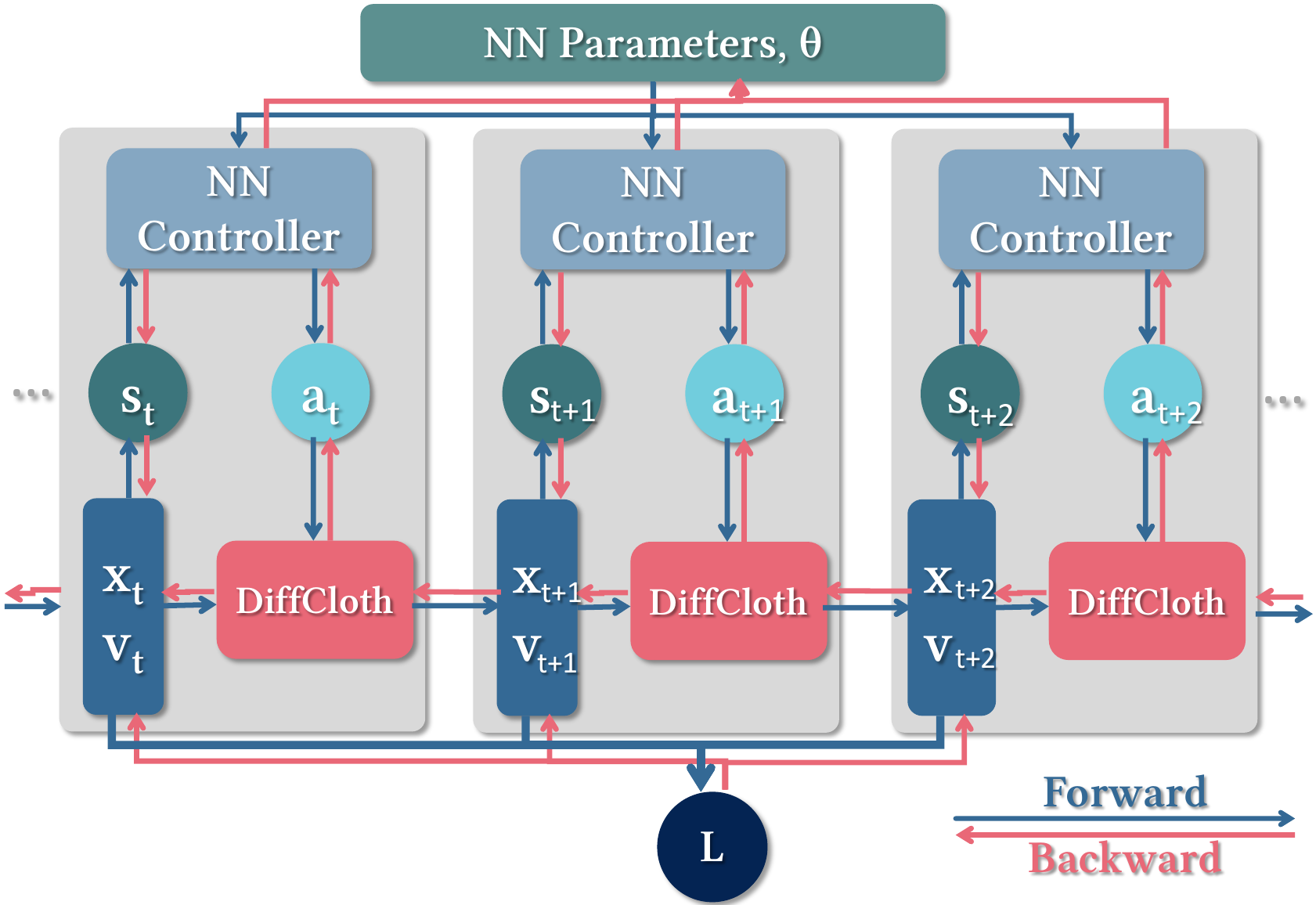}
    \caption{\hlred{We present the computation graph of the "Hat Controller" task in Sec.~\ref{sec:hat_controller}. We embed the neural network policy into our differentiable simulation pipeline. At time step $t$ during forward simulation (blue arrows), the current particle states $(\x_t,\vel_t)$ are transformed into a state vector $\mathbf{s}_t$, which is input to the neural network policy $\pi_\theta$ to produce an action vector $\mathbf{a}_t = \pi_\theta(\mathbf{s}_t)$. DiffCloth then simulates a new state  $(\x_{t+1},\vel_{t+1})$ using current particle state and the generated action vector. The whole simulation sequence $\{x_i,v_i\}$ is used to compute a loss. During backward propagation (red arrows), the above computation pass is reversed. }}
    \label{fig:hatController_computationgraph}
\end{figure}

\hlred{
\subsection{Hat Controller}
\label{sec:hat_controller}
We end this section with an advanced ``Hat'' task. 
Unlike the previous open-loop trajectory optimization with a fixed starting position, the goal of this new task is to train a generalizable closed-loop controller that can put on the hat from a \emph{random} starting position sampled from a fixed-radius hemisphere around the head. Specifically, we train a closed-loop control policy $\mathbf{a}_t = \pi_\theta(\mathbf{s}_t)$, which takes as input the current state $\mathbf{s}_t$ of the task and outputs an action vector $\textbf{a}_t$ at each time step $t$. The state vector $\mathbf{s}_t$ includes the hat node positions, the orientation of the hat, and the distance between the two end effectors, and the action vector $\mathbf{a}_t$ represents the position of the two end effectors at the next time step. We represent the control policy $\pi_\theta$ as a neural network parametrized by $\theta$ consisting of two fully-connected hidden layers with 64 neurons and \texttt{tanh} activation functions (117126 parameters in total). To train the controller with gradient-based optimization, we integrate the neural network policy with our differentiable simulator as described in Fig.~\ref{fig:hatController_computationgraph}.}

\hlred{In each epoch during training, we randomly sample 20 starting positions of the hat and compute a loss averaged from all simulation sequences. The loss function of each sequence is defined by $L=L_{\text{deform}}+L_{\text{target}}+L_{\text{dir}}$, where $L_{\text{deform}}$ measures the stretching of the hat using the distance change between the two end-effectors, $L_{\text{target}}$ 
measures the L2-distance between the last-time-step pose of the hat and the target pose, and $L_{\text{dir}}$ is the orientation difference between the last-time-step pose and the target pose of the hat. The gradient of the loss is then computed by our differentiable simulation framework and used by a gradient-based optimizer (Adam~\cite{kingma2017adam}) to update the policy parameters. For testing, we evaluate the controller from 20 fixed starting position configurations uniformly sampled on the hemisphere surface.}

\hlred{
To compare our method with gradient-free methods, we also solve the task with Reinforcement Learning (RL), which has been widely used to train complex neural network policies for robot-assisted dressing problems~\cite{Clegg2018LearningtoDress}. Specifically, we compare our gradient-based method with PPO~\cite{PPO}, a state-of-the-art RL algorithm. We similarly sample a starting position from the hemisphere in each iteration when training PPO. For a fair comparison, we design the reward function $r_t$ to be the sum of the negative counterpart of $L$ plus a constant to avoid negative rewards, and observation of the environment to be $\mathbf{s}_t$. We evaluate Adam and PPO using $L$ as the common metric and plot the optimization curve at the bottom of Fig.~\ref{fig:hatController_sequences}. We see that both methods reach a similar final loss, but with our differentiable simulation framework, the gradient-based method reaches its final loss with a 85x speedup (Adam uses 23200 time steps; PPO uses 1978000 time steps).
}

\hlred{
After the training process converges, both our gradient-based method and PPO successfully move the hat onto the head from all 20 testing positions.  We visualize the trajectories generated by Adam's trained policy from 10 testing starting positions at the top of Fig.~\ref{fig:hatController_sequences}. Unlike existing differentiable simulation papers~\cite{Du21_DiffPD,Hu19_ChainQueen,hu2019difftaichi,Liang19_DiffCloth} that train a closed-loop network controller only for a fixed state, we highlight that we use differentiable simulation to train the network from multiple, random states and study its generalizability in a test set of unseen states. This allows us to conduct a fairer comparison between gradient-based optimization method and RL methods, which is typically overlooked in previous papers.}

\section{Conclusions, Limitation, and Future Work}\label{sec:conclusions}

In this paper, we presented a differentiable cloth simulator built on PD with Signorini-Coulomb frictional contact. Our differentiable simulator is different from existing papers in its simultaneous accommodation of rich and frequent (self-)contact, Signorini-Coulomb contact law, and differentiability in cloth simulation. \hlred{We analyzed the numerical properties of gradients from our differentiable simulator, including the source of discontinuities and its empirical speedup over gradient-free approaches in high-dimensional problems.} We additionally presented an iterative solver that exploits the contact gradients to speed up the backpropagation and observed a substantial speedup (up to an order of magnitude with low-precision simulation) over direct solvers. Our differentiable cloth simulator enabled gradient-based optimization methods in a diverse set of applications, for which traditional gradient-free methods are generally much less sampling efficient. \hlred{In particular, we presented a preliminary study on training a generalizable closed-loop controller using differentiable simulation, in which our approach and PPO achieved comparable performance and generalizability, but we used much less time.}

There are still quite a few limitations in our method that are worth further investigation. First, since our method is built on PD, it also limits the choice of material models. It would be useful to generalize the current framework and support more physically accurate cloth material models, e.g., the piecewise linear elastic model described in~\citet{Wang2011DataDriven}.

Second, the contact model we use from~\citet{Ly20_PDdryFriction} does not take into account vertex-face or edge-edge self-collisions. While we empirically observed that handling only vertex-vertex self-collisions managed to produce plausible results with medium mesh resolution, vertex-face and edge-edge collision detection and handling is still highly desirable for a more physically realistic cloth simulator.

\hlred{Third, although we have identified some possible sources of discontinuities and non-smoothness in our differentiable simulator with empirical experiments, their effects on gradient-based optimization still require a thorough investigation. In particular, the locally bumpy energy landscape we observed in Fig.~\ref{fig:evaluation_collision_num} due to changes in contact sets makes us question both the necessity and the usefulness of exact gradients in optimization, although Sec.~\ref{sec:applications} implies that gradients were still helpful in many downstream applications. Noting that the bumpiness in Fig.~\ref{fig:evaluation_collision_num} is local and the global view of the energy landscape is still smooth, we hypothesize an inexact but smoothed gradients would be more powerful, which we leave as future work.}

Fourth, there is no theoretical guarantee on the convergence of the iterative solver we implemented in backpropagation. Although non-convergence is uncommon in our experiments, it introduces a costly switch to the slower direct solver, which we hope to fully resolve in future work.

Finally, many of our applications were in simulation only. It would be more exciting if these results could be replicated in real-world settings. We consider connecting our differentiable cloth simulator to more real-world applications, including real-world robot-assisted dressing, material parameter identification for real-world fabric samples, computational design for sports suits, and so on. Closing the sim-to-real gap is a nontrivial problem, in which we believe our differentiable simulator could play a beneficial role.

\begin{acks}
We thank Marco Renedo for his helpful discussions on the preconditioners, Junbang Liang for his help with running the baseline comparison code, and the anonymous reviewers for their helpful comments. This work was supported in part by the Defense Advanced Research Projects Agency (DARPA) under grant No. FA8750-20-C-0075.

\end{acks}

\bibliographystyle{ACM-Reference-Format}
\bibliography{bibliography}
\appendix

\section{Experiment Run Time}\label{sec:appendixExpRuntime}
We run all optimizations on a workstation of 80 CPU cores and 80G memory. Depending on the problem complexity, the wallclock time of running these optimizations varies from less than 30 minutes to 2 hours for all methods. We report the run time for optimizing the examples shown in Sec.~\ref{sec:applications} using our gradient-based method in Table~\ref{tab:appTime}. 

\begin{table}[htb]
\caption{Run time for our gradient-based optimization. We report the average wall-clock time for optimizing the examples shown in Sec.~\ref{sec:applications}. We recall the simulation complexity of each example by reiterating its total Degrees of Freedom (``Dof''), number of time steps in each forward simulation sequence and back-propagation (``Time Steps'') and time interval (``$h[s]$''). ``Fwd.[s]"" and ``Back.[s]"" report the mean wall-clock time for performing one iteration of forward simulation and back-propagation respectively, averaged across all optimization iterations for all initial seeds.}
\label{tab:appTime}
\begin{tabular}{c|c|c|c|c|c}
\toprule
\multicolumn{1}{l|}{\textbf{Name}} & \multicolumn{1}{c|}{\textbf{Dof}} & \multicolumn{1}{c|}{\textbf{Time Steps}} & \multicolumn{1}{c|}{\textbf{$h$ {[}s{]}}} & \multicolumn{1}{l|}{\textbf{Fwd.{[}s{]}}} & \multicolumn{1}{l}{\textbf{Back.{[}s{]}}} \\ \midrule
T-shirt& 4278& 250& 1/90& 141.5& 13.2\\
Sphere& 1875& 350& 1/180& 5.1& 4.2\\
Hat& 1737& 400& 1/100& 57.9& 20.7\\
Sock& 3165& 400& 1/160& 89.9& 14.3\\
Dress& 10902& 125& 1/120& 266.3& 84.7\\
Flag& 1620& 100& 1/120& 13.7& 1.6 \\ \bottomrule
\end{tabular}
\end{table}

\hlred{
\section{Experiment Details}\label{sec:appendixExpDetails}
We provide detailed information for the examples shown in Sec.~\ref{sec:applications}, including their setup, the exact form of their loss function, and their decision variables in optimization.
\subsection{System Identification}
\paragraph{T-shirt} %
The loss function is defined as
\begin{align}
L = \sum_{n=1}^N \norm{\x^{\text{current}}_n - \x^{\text{target}}_n }_2,
\end{align}
where $N=240$ is the total number of time steps, $\x^{\text{current}}_n$ the position of the mesh during optimization, and $\x^{\text{target}}_n$ the position of the ground-truth simulation generated using our system.  There are 4 parameters (6 DoFs) to be optimized: $\theta_{\text{T-shirt}} = (k_\text{stretch}, \phi, \omega, \dd)$,  where $k_\text{stretch}$ controls the stretching stiffness of the cloth material and $(\phi, \omega, \dd)$ controls the parameterized external wind force: each node receives a three-dimensional wind force $0.5[\sin(\omega t + \phi) + 1.0]\dd$.}

\hlred{
\paragraph{Sphere} The loss function is defined the same as in the ``T-shirt'' example above except that the total time $N=300$. The decision variable is the 1-DoF frictional coefficient of the sphere.}

\hlred{
\subsection{Robot-Assisted Dressing}
\paragraph{Hat} %
The loss function is defined as the L2 norm between the last-time-step position of the hat $\x^{\text{current}}_N$ and a target position $\x^{\text{target}}_N$ generated by translating the initial pose of the hat onto the head:
\begin{align}
    L = \norm{\x^{\text{current}}_N - \x^{\text{current}}_N}_2,
\end{align} where $N=400$ is the total number of time steps. The trajectory of each of the two end effectors is controlled by a cubic Hermite spline. Each spline has $3$ parameters (9 DoFs): $\theta_{\text{spline}}=(\tb_1, \tb_2, \p_{\text{end}})$, where $\tb_1$ and $\tb_2$ are the two tangents of the spline and $\p_{\text{end}}$ is the endpoint of the spline.}

\hlred{\paragraph{Sock} The goal is to optimize for the trajectory of the three end effectors holding onto a sock so that the sock can be put onto a synthetic foot model from an initial starting position. To guide the end effectors to first hook the opening of the sock onto the tip of the foot, then slide the sock upward onto the leg, the loss function is designed as
\begin{align}
L = \sum_{n\in\{\frac{N}{2}, N\}} \sum_{(p_{\text{foot}}, p_{\text{sock}}) \in \mathcal{P}_n} \norm{ \x^{\text{target}}_{N,p_{\text{foot}}} - \x^{\text{current}}_{n,p_{\text{sock}}}}_2,
\end{align}
where $\mathcal{P}_t$ defines a set of manually selected keypoint correspondence pairs $(p_{\text{foot}}, p_{\text{sock}})$ between the sock mesh $\x_n^{\text{current}}$ and the foot model  $\x_N^{\text{target}}$ at halfway $t=\frac{N}{2}$ and the end of the simulation $t=N$, respectively. We sum up the L-2 norm of the position difference between each correspondence.  The optimization parameters are the Hermite spline parameters for each of the four end effectors defined similarly as in the "Hat" example (36 DoFs in total).
}

\hlred{
\subsection{Inverse Design}
\paragraph{Dress} %
The loss function is defined as
\begin{align}
L = \sum_{p\in \mathcal{P}} (h_p- h)^2,
\end{align}
where $h$ is the calculated target height of the bottom of the dress when the desired apex angle is reached.  $\mathcal{P}$ is the set of all points located at the bottom of the dress, and $h_p$ is the height of the point  $p$ (corresponding to the y coordinate of the point in our implementation). There are 2 parameters (2 DoFs) that are being optimized: $\theta_{\text{Dress}} = (d, k_{\text{bend}})$ where $d$ is the density of the fabric and $k_{\text{bend}}$ is the bending stiffness of the fabric.
}

\hlred{
\subsection{Real-to-Sim Example}
In this task, we optimize for the material parameters of the flag and the parameters of a simple wind model to match the motion trajectory of a flag to real-world captured data. The wind model is defined as
\begin{align}
\fex = \frac{\sin(\omega t + \phi) + 1.0}{2} \mathbf{k}  \dd^\top,
\end{align}
where $\dd$ is a 3D vector to be optimized and  $\mathbf{k} \in \R^{m}$ a constant coefficient vector with each entry equal to a node's inverse distance to the flag center evaluated at the first time step. There are 8 parameters to be optimized $\theta_{\text{flag}} =  (k_\text{stretch}, k_\text{bend}, \rho, \phi, \omega, \dd)$, where $k_\text{stretch}$ and $k_\text{bend}$ are the stretching and bending stiffness of the fabric, $\rho$ the density of the fabric, and $\phi, \omega, \dd$ the parameters of the wind model defined above.
}

\hlred{
\subsection{Hat Controller}
In this task, the goal is to optimize for the neural network parameters of the hat controller so that the two end effectors put a hat on a head model from any starting position defined on a fixed-radius hemisphere centered at the head model. During training, we uniformly sample 20 starting positions on the hemisphere for each epoch, and compute the loss averaged from all simulation sequences. We define the loss function as
\begin{align}
    L=L_{\text{deform}}+L_{\text{target}}+L_{\text{dir}}.
\end{align}
$L_{\text{deform}}$ minimizes the distance change between the two end effectors and is defined as $L_{\text{deform}} =\sum_{n=1}^N \norm{\x_{n,e_1} - \x_{n,e_2} }_2 $ where $e_1$ and $e_2$ are the indices of the nodes pulled by the two end effectors.  $L_{\text{target}}$ minimizes the L2-distance between the poses of the hat and the target pose at the last few frames (20 in our implementation) and is defined as
\begin{align}
L_{\text{target}}= \sum_{n=N-20}^N \norm{\x_n^{\text{current}} - \x_n^{\text{target}} }_2.
\end{align}
$L_{\text{dir}}$ minimizes the orientation difference between the last-time-step pose and the target pose of the hat and is defined as
\begin{align}
L_{\text{dir}} = \textbf{d}_{\text{target}}  \cdot  \textbf{d}_{\text{current}}
\end{align}
where $\textbf{d}_{\text{target}}$ is the up vector of the hat at the target pose and $ \textbf{d}_{\text{current}}$ is defined similarly to the hat at the last time step. 
}

\section{Optimization Results For All Random Seeds}\label{sec:appendixAllSeedsExp}
In this section we report the optimization results for all random seeds in Table~\ref{tab:appSeed1} and Table~\ref{tab:appSeed2}. For most experiments, L-BFGS-B achieves lower or comparable optimized loss than the gradient-free methods using a fraction (often to an order of magnitude) of time steps, thanks to the gradient information provided by our differentiable simulator. For some random seeds, L-BFGS-B converges to a relatively large final loss percentage, possibly due to being stuck in a local minimum. In practice, it is common and recommended to run gradient-based optimizations with several initial seeds to alleviate this problem, and is the rationale behind why we choose to report the minimum loss achieved across all random seeds in the results shown in Table~\ref{tab:appLoss} and plot the minimum loss envelop in Fig~\ref{fig:high_dim_opt},~\ref{fig:hat},~\ref{fig:sock},~\ref{fig:flag}. It is also worth mentioning that the examples shown Sec.~\ref{sec:applications} have a relative low number of design variables, while the speed-up of gradient-based methods becomes more evident with more design variables as demonstrated by the ``Flying Napkin'' experiment in Fig.~\ref{fig:high_dim_opt}. 
\begin{table*}[bp]
\caption{We report the optimization results for all random seeds of all methods in the ``T-shirt'' and ``Flag'' examples. For each random seed, we report its initial loss and final loss achieved by each optimization method. ``Final Loss Percentage (\%)'' reports the optimized loss as a \textit{percentage (0-100\%)} of the initial loss. ``Convergence Time Steps'' reports the number of time steps used by each method to reach its final loss. For all tasks, we also summarize the minimum, average and median statistics across all random seeds for each column. For each metric (``Final Loss'', ``Final Loss Percentage (\%)'', ``Convergence Time Steps'') and each row, the minimum number across the three optimization methods is marked in bold. }
\label{tab:appSeed1}
\begin{tabular}{c|c|ccc|ccc|ccc}
\toprule
 &  & \multicolumn{3}{c|}{\textbf{Final Loss}}   & \multicolumn{3}{c|}{\textbf{Final Loss Percentage (\%)}} & \multicolumn{3}{c}{\textbf{Convergence Time Steps}} \\ \cline{3-5}\cline{6-8}\cline{9-11}
    &   {\multirow{-2}{*}{\textbf{Initial Loss}}}   & \textbf{L-BFGS-B} & \textbf{CMA-ES} & \textbf{(1+1)-ES} & \textbf{L-BFGS-B}  & \textbf{CMA-ES}  & \textbf{(1+1)-ES} & \textbf{L-BFGS-B}  & \textbf{CMA-ES}  & \textbf{(1+1)-ES}  \\ \toprule
\multirow{8}{*}{\textbf{T-shirt}} & 22.49   & \textbf{0.042} & 0.052   & 1.416   & \textbf{0.2} & 0.2   & 6.2   & \textbf{2500} & 41750 & 42000 \\
   & 60.00   & \textbf{0.079} & 0.269   & 0.248   & \textbf{0.1} & 0.4   & 0.4   & \textbf{7250} & 46750 & 9750  \\
   & 6.62    & 6.560   & 0.169   & \textbf{0.115} & 99.0         & 2.7   & \textbf{1.8} & \textbf{3250} & 17500 & 47750 \\
   & 30.93   & \textbf{0.035} & 0.286   & 0.213   & \textbf{0.1} & 0.9   & 0.7   & \textbf{2250} & 41000 & 36750 \\
   & 10.32   & \textbf{0.035} & 0.078   & 0.159   & \textbf{0.3} & 0.8   & 1.6   & \textbf{6250} & 27000 & 19750  \\ \cmidrule{2-11} 
\textbf{}    & \textbf{MIN}     & \textbf{0.035} & 0.052   & 0.115   & \textbf{0.1} & 0.2   & 0.4   & \textbf{2250} & 17500 & 9750  \\
   & \textbf{AVERAGE} & 1.350   & \textbf{0.171} & 0.430   & 20.0         & \textbf{1.0} & 2.1   & \textbf{4300} & 34800 & 31200 \\
   & \textbf{MEDIAN}  & \textbf{0.042} & 0.169   & 0.213   & \textbf{0.2} & 0.8   & 1.6   & \textbf{3250} & 41000 & 36750   \\  \midrule
\multirow{13}{*}{\textbf{Flag}} & 2.88    & \textbf{0.137} & 1.118   & 1.154   & \textbf{4.8}  & 38.9   & 40.1         & \textbf{3400} & 17300         & 24800 \\
     & 3.84    & 1.136   & \textbf{0.945} & 1.079   & 29.6   & \textbf{24.6} & 28.1         & 1100   & \textbf{900}  & 27600 \\
     & 4.07    & 1.175   & \textbf{0.304} & 1.021   & 28.9   & \textbf{7.5}  & 25.1         & 5200   & \textbf{1200} & 28500 \\
     & 5.08    & \textbf{0.595} & 0.958   & 1.043   & \textbf{11.7} & 18.9   & 20.5         & \textbf{3800} & 4500   & 26600 \\
     & 4.08    & 1.980   & \textbf{0.997} & 1.053   & 48.6   & \textbf{24.5} & 25.8         & \textbf{200}  & 29200         & 27600 \\
     & 3.26    & \textbf{0.190} & 0.743   & 0.952   & \textbf{5.8}  & 22.8   & 29.2         & \textbf{4500} & 27500         & 28400 \\
     & 3.59    & 0.863   & \textbf{0.697} & 1.134   & 24.0   & \textbf{19.4} & 31.5         & \textbf{800}  & 9900   & 28100 \\
     & 3.68    & \textbf{0.156} & 0.711   & 1.089   & \textbf{4.2}  & 19.3   & 29.6         & \textbf{9800} & 22500         & 24600 \\
     & 3.73    & 1.095   & \textbf{1.077} & 1.109   & 29.4   & \textbf{28.9} & 29.8         & \textbf{3100} & 12600         & 28900 \\
     & 9.98    & 0.861   & 1.086   & \textbf{0.715} & 8.6    & 10.9   & \textbf{7.2} & \textbf{1800} & 5400   & 9200     \\  \cmidrule{2-11} 
    & \textbf{MIN}     & \textbf{0.137} & 0.304   & 0.715   & \textbf{4.2}  & 7.5    & 7.2   & \textbf{200}  & 900    & 9200  \\
     & \textbf{AVERAGE} & \textbf{0.819} & 0.864   & 1.035   & \textbf{19.6} & 21.6   & 26.7         & \textbf{3370} & 13100         & 25430 \\
     & \textbf{MEDIAN}  & \textbf{0.862} & 0.952   & 1.066   & \textbf{17.9} & 21.1   & 28.7         & \textbf{3250} & 11250         & 27600 \\ \bottomrule
\end{tabular}
\end{table*}
\begin{table*}[htb]
\caption{Similar table as Table~\ref{tab:appSeed1} for the ``Hat'', ``Sock'', ``Sphere'' and ``Dress'' examples.  }
\label{tab:appSeed2}
\begin{tabular}{c|c|ccc|ccc|ccc}
\toprule
 &  & \multicolumn{3}{c|}{\textbf{Final Loss}}   & \multicolumn{3}{c|}{\textbf{Final Loss Percentage (\%)}} & \multicolumn{3}{c}{\textbf{Convergence Time Steps}} \\ \cline{3-5}\cline{6-8}\cline{9-11}
    &   {\multirow{-2}{*}{\textbf{Initial Loss}}}   & \textbf{L-BFGS-B} & \textbf{CMA-ES} & \textbf{(1+1)-ES} & \textbf{L-BFGS-B}  & \textbf{CMA-ES}  & \textbf{(1+1)-ES} & \textbf{L-BFGS-B}  & \textbf{CMA-ES}  & \textbf{(1+1)-ES}  \\ \toprule
\multirow{8}{*}{\textbf{Hat}} & 54.60   & 10.391         & 0.754 & \textbf{0.330} & 19.0         & 1.4 & \textbf{0.6} & \textbf{4400} & 47600 & 50000 \\
   & 33.81   & 0.402   & 0.665 & \textbf{0.105} & 1.2   & 1.9 & \textbf{0.3} & \textbf{2000} & 46000 & 36400 \\
   & 43.45   & \textbf{0.091} & 0.723 & 0.236   & \textbf{0.2} & 1.6 & 0.5   & \textbf{5600} & 42800 & 44000 \\
   & 48.63   & \textbf{0.105} & 0.283 & 1.290   & \textbf{0.2} & 0.6 & 2.6   & \textbf{4000} & 48000 & 45600 \\
   & 21.83   & \textbf{0.096} & 0.946 & 0.314   & \textbf{0.4} & 4.3 & 1.4   & \textbf{2800} & 38800 & 30000   \\ \cmidrule{2-11} 
      & \textbf{MIN}     & \textbf{0.091} & 0.283 & 0.105   & \textbf{0.2} & 0.6 & 0.3   & \textbf{2000} & 38800 & 30000 \\
   & \textbf{AVERAGE} & 2.217   & 0.674 & \textbf{0.455} & 4.2   & 2.0 & \textbf{1.1} & \textbf{3760} & 44640 & 41200 \\
   & \textbf{MEDIAN}  & \textbf{0.105} & 0.723 & 0.314   & \textbf{0.4} & 1.6 & 0.6   & \textbf{4000} & 46000 & 44000 \\ \midrule
\multirow{8}{*}{\textbf{Sock}} & 46.02   & \textbf{1.980} & 7.267  & 5.335   & \textbf{4.3}  & 15.8 & 11.6         & \textbf{8800}  & 42800 & 47600 \\
    & 41.22   & \textbf{3.243} & 6.131  & 8.396   & \textbf{7.9}  & 14.9 & 20.4         & \textbf{16000} & 44000 & 32400 \\
    & 39.10   & 8.856   & 10.547 & \textbf{3.047} & 22.6   & 27.0 & \textbf{7.8} & \textbf{4400}  & 38400 & 46000 \\
    & 17.08   & \textbf{2.589} & 4.149  & 7.830   & \textbf{15.2} & 24.3 & 45.8         & \textbf{15200} & 31200 & 24000 \\
    & 33.29   & \textbf{2.652} & 4.126  & 5.791   & \textbf{8.0}  & 12.4 & 17.4         & \textbf{5200}  & 49200 & 48000    \\\cmidrule{2-11} 
         & \textbf{MIN}     & \textbf{1.980} & 4.126  & 3.047   & \textbf{4.3}  & 12.4 & 7.8   & \textbf{4400}  & 31200 & 24000 \\
    & \textbf{AVERAGE} & \textbf{3.864} & 6.444  & 6.080   & \textbf{11.6} & 18.9 & 20.6         & \textbf{9920}  & 41120 & 39600 \\
    & \textbf{MEDIAN}  & \textbf{2.652} & 6.131  & 5.791   & \textbf{8.0}  & 15.8 & 17.4         & \textbf{8800}  & 42800 & 46000 \\ \midrule
\multirow{13}{*}{\textbf{Sphere}} & 0.46    & 0.002 & \textbf{0.000} & 0.003   & 0.4   & 0.0   & 0.6   & \textbf{2450} & 18900         & 11200 \\
   & 0.90    & 0.064 & \textbf{0.000} & \textbf{0.000} & 7.2   & \textbf{0.0} & \textbf{0.0} & \textbf{3500} & 36050         & 5600  \\
   & 0.58    & 0.002 & \textbf{0.000} & \textbf{0.000} & 0.3   & \textbf{0.0} & \textbf{0.0} & \textbf{1400} & 24850         & 10150 \\
   & 2.20    & 0.904 & \textbf{0.000} & \textbf{0.000} & 41.1  & \textbf{0.0} & \textbf{0.0} & \textbf{700}  & 31850         & 8050  \\
   & 4.03    & 0.904 & \textbf{0.000} & 0.031   & 22.4  & \textbf{0.0} & 0.8   & \textbf{700}  & 15050         & 26250 \\
   & 0.90    & 0.903 & \textbf{0.000} & 0.020   & 100.0 & \textbf{0.0} & 2.0   & \textbf{350}  & 26600         & 44100 \\
   & 0.90    & 0.002 & 0.002   & \textbf{0.000} & 0.2   & 0.2   & \textbf{0.0} & 6650   & \textbf{5600} & 18200 \\
   & 0.89    & 0.893 & \textbf{0.000} & 0.008   & 100.0 & \textbf{0.0} & 0.8   & \textbf{350}  & 48300         & 46550 \\
   & 0.88    & 0.861 & \textbf{0.000} & 0.001   & 97.6  & \textbf{0.0} & 0.1   & \textbf{1400} & 49700         & 9100  \\
   & 0.85    & 0.514 & \textbf{0.000} & \textbf{0.000} & 60.4  & \textbf{0.0} & \textbf{0.0} & \textbf{3150} & 35700         & 15400    \\ \cmidrule{2-11} 
  & \textbf{MIN}     & 0.002 & \textbf{0.000} & \textbf{0.000} & 0.2   & \textbf{0.0} & \textbf{0.0} & \textbf{350}  & 5600   & 5600  \\
   & \textbf{AVERAGE} & 0.505 & \textbf{0.000} & 0.006   & 43.0  & \textbf{0.0} & 0.4   & \textbf{2065} & 29260         & 19460 \\
   & \textbf{MEDIAN}  & 0.688 & \textbf{0.000} & 0.001   & 31.7  & \textbf{0.0} & 0.1   & \textbf{1400} & 29225         & 13300 \\ \midrule
\multirow{15}{*}{\textbf{Dress}} & 2.41    & 1.820   & \textbf{0.712} & 0.845   & 75.4   & \textbf{33.3} & 39.6   & \textbf{375}  & 3125  & 12750   \\
   & 1.35    & 0.716   & 0.830   & \textbf{0.696} & 53.2   & 60.4   & \textbf{50.6} & \textbf{1125} & 49875 & 10000   \\
   & 1.57    & 1.406   & 0.824   & \textbf{0.782} & 89.3   & 52.3   & \textbf{49.6} & \textbf{1625} & 31875 & 9750    \\
   & 1.03    & 0.841   & 0.825   & \textbf{0.822} & 82.0   & 80.3   & \textbf{80.0} & \textbf{1625} & 29375 & 19750   \\
   & 0.90    & 0.880   & 0.824   & \textbf{0.823} & 97.7   & 90.3   & \textbf{90.1} & \textbf{750}  & 19750 & 16250   \\
   & 1.49    & 1.490   & 0.832   & \textbf{0.698} & 99.9   & 55.4   & \textbf{46.5} & \textbf{875}  & 44000 & 20750   \\
   & 1.09    & 0.875   & \textbf{0.828} & 0.830   & 80.3   & \textbf{75.2} & 75.4   & \textbf{250}  & 35000 & 17250   \\
   & 1.88    & \textbf{0.693} & 0.861   & 0.792   & \textbf{37.0} & 45.4   & 41.8   & \textbf{1375} & 25875 & 40625   \\
   & 1.83    & 1.305   & \textbf{0.823} & 0.858   & 71.2   & \textbf{56.1} & 58.5   & \textbf{1125} & 50000 & 10250   \\
   & 1.27    & 1.219   & \textbf{0.824} & 0.854   & 96.1   & \textbf{64.4} & 66.8   & \textbf{500}  & 49500 & 14000   \\
   & 1.31    & 1.178   & \textbf{0.814} & 0.872   & 90.1   & \textbf{62.2} & 66.6   & \textbf{2375} & 4125  & 10250   \\
   & 1.26    & 0.856   & 0.824   & \textbf{0.823} & 68.2   & 65.0   & \textbf{65.0} & \textbf{1000} & 47125 & 11000  \\ \cmidrule{2-11} 
       & \textbf{MIN}     & \textbf{0.693} & 0.712   & 0.696   & 37.0   & \textbf{33.3} & 39.6   & \textbf{250}  & 3125  & 9750  \\
   & \textbf{AVERAGE} & 1.107   & 0.818   & \textbf{0.808} & 78.4   & 61.7   & \textbf{60.9} & \textbf{1083} & 32469 & 16052 \\
   & \textbf{MEDIAN}  & 1.029   & 0.824   & \textbf{0.823} & 81.1   & \textbf{61.3} & 61.7   & \textbf{1063} & 33438 & 13375 \\ \bottomrule
\end{tabular}
\end{table*}

\end{document}